\documentclass{article}
\usepackage[utf8]{inputenc}
\usepackage{authblk}
\usepackage{siunitx}
\usepackage{caption}
\usepackage{graphicx}
\usepackage{subfigure}
\usepackage{float}
\usepackage{amsmath}
\usepackage{amsthm}
\usepackage{array}
\usepackage{hyperref}
\usepackage{cleveref}
\usepackage{appendix}
\usepackage[greek, ngerman, english]{babel}
\usepackage{multirow}
\usepackage{xr}
\usepackage{lineno}

\newcommand{\mpmin}[1]{\underset{#1}{\text{min}}\,}
\newcommand{\mpst}{\text{s.t.}\,}
\newcommand{\mple}{{}\le}

\begin{document}
	
	\title{Data-driven Product-Process Optimization of N-isopropylacrylamide Microgel Flow-Synthesis}
	\author[1]{Luise F. Kaven}
	\author[2]{Artur M. Schweidtmann}
	\author[1]{Jan Keil}
	\author[1]{Jana Israel}
	\author[3,4]{Nadja Wolter}
	\author[1,5,*]{Alexander Mitsos}
	
	\affil[1]{Process Systems Engineering (AVT.SVT), RWTH Aachen University, Forckenbeckstr. 51, 52074 Aachen, Germany}
	\affil[2]{Department of Chemical Engineering, Delft University of Technology, Van der Maasweg 9, 2629 HZ Delft, The Netherlands}
	\affil[3]{DWI - Leibniz Institute for Interactive Materials e.V., Forckenbeckstr. 50, 52074 Aachen, Germany}
	\affil[4]{Functional and Interactive Polymers, Institute of Technical and Macromolecular Chemistry, RWTH Aachen University, Forckenbeckstr. 50, 52074 Aachen, Germany}
	\affil[5]{JARA-soft, RWTH Aachen University, Templergraben 55, 52056 Aachen, Germany}
	\affil[*]{Corresponding author, E-mail: amitsos@alum.mit.edu, Forckenbeckstr. 51, 52074 Aachen, Germany}
	
	\date{June 2023}

	\maketitle
	
	\abstract{Microgels are cross-linked, colloidal polymer networks with great potential for stimuli-response release in drug-delivery applications, as their size in the nanometer range allows them to pass human cell boundaries.
		For applications with specified requirements regarding size, producing tailored microgels in a continuous flow reactor is advantageous because the microgel properties can be controlled tightly.
		However, no fully-specified mechanistic models are available for continuous microgel synthesis, as the physical properties of the included components are only studied partly.
		To address this gap and accelerate tailor-made microgel development, we propose a data-driven optimization in a hardware-in-the-loop approach to efficiently synthesize microgels with defined sizes.
		We optimize the synthesis regarding conflicting objectives (maximum production efficiency, minimum energy consumption, and the desired microgel radius) by applying Bayesian optimization via the solver ``Thompson sampling efficient multi-objective optimization'' (TS-EMO).
		We validate the optimization using the deterministic global solver ``McCormick-based Algorithm for mixed-integer Nonlinear Global Optimization'' (MAiNGO) and verify three computed Pareto optimal solutions via experiments.
		The proposed framework can be applied to other desired microgel properties and reactor setups and has the potential of efficient development by minimizing number of experiments and modelling effort needed.}
	
	\vspace*{5mm}
	
	
	\section{Introduction} \label{sec:int}
	The size of microgels in the nano- and micrometer range and the microgel's ability to react reversibly to external stimuli of temperature, pH, or electrical potential in the surrounding medium~\cite{Pich.2011} is highly relevant for their application.
	The microgel size has been studied for biomedical~\cite{Keskin.2019, Switacz.2020, Zhang.2019}, phase separation~\cite{Faulde.2018, Destribats.2014, Richtering.2012}, and catalysis~\cite{Khan.2020} applications.
	Microgels in the nanometer size range have previously been applied for biomedical purposes, e.g., for drug delivery agents for medical uptake and release~\cite{Switacz.2020, Zhang.2019} or implant coating~\cite{Keskin.2019}.
	In biomedical applications, microgels are particularly relevant, as their small size allows them to pass the human cell boundary~\cite{Switacz.2020}.
	For the cellular uptake, it was found that microgels of a hydrodynamic radius in the swollen state (at \SI{20}{\degreeCelsius}) above \SI{400}{\nano\meter} and a cross-linker content above \SI{10}{\mol\percent} prevent microgel internalization.

	The synthesis of microgels in flow reactors can overcome shortcomings of batch reactors, e.g., limited production capacity and downtime between batches, and enhances product development, intensifies production, and facilitates reaction scale-up~\cite{Wolff.2018, Kather.2018, Fandrich.2020, Kaven.2021}.
	Furthermore, including process analytical technology in flow reactors allows in-line monitoring and process control under highly reproducible conditions~\cite{Fandrich.2020, Kaven.2021, Fandrich.2023}.
	Thus, continuous production enables the reliable synthesis of microgels.

	To unfold the full potential of microgels, accelerating the development of tailor-made microgels is desirable.
	A faster development can be achieved by producing microgels in a continuous reactor mode, as it simplifies up-scaling to large-scale industrial production.
	Furthermore, model-based approaches facilitate the optimization of microgels with tailored properties.
	Computational models for describing microgel growth during the synthesis are very sparse and comprise mechanistic models suited for batch reaction exclusively~\cite{Janssen.2019, Jung.2019, Hoare.2006, Janssen.2017, Janssen.2018, Jung.2019b}.
	Our previous study~\cite{Kaven.2021} revealed significant deviations between the reaction progress in batch and flow reactors in the microgel synthesis.
	In particular, we cannot transfer the batch model equations straight to a plug-flow model, but rather we must consider diffusion effects, temperature distribution, and rheological aspects.
	The physical properties such as diffusivity coefficient and viscosity are not known during the microgel synthesis, so mechanistic modeling of the flow process is restricted.
	
	To address this gap, we propose a data-driven hardware-in-the-loop optimization for N-isopropylacrylamide-based microgels, one of the most widely studied thermo-responsive microgel systems.
	The data-driven approach facilitates the reaction optimization of the microgel synthesis in flow.
	We apply Thompson sampling efficient multi-objective optimization~(TS-EMO)~\cite{Bradford.2018} to enhance the experimental synthesis design iteratively.
	The TS-EMO solver is based on the Thompson sampling algorithm, a popular approach in Bayesian optimization.
	Bayesian optimization searches for a (global) optimum with a focus on efficiency, i.e., aiming for small number of function evaluations.
	Efficiency is crucial when function evaluations are costly, e.g., require experimentation or extensive computation.
	In Bayesian optimization, a probabilistic model (also called surrogate or digital twin) of the objective function is constructed and iteratively updated as new data points are evaluated.
	The surrogate models are constructed via Gaussian Processes~(GPs). 
	GPs are considered an effective surrogate model as they provide predictions and variance estimates while relying on relatively few data points~\cite{Shields.2021}.
	Black-box optimization involving GPs for chemical synthesis has been successfully applied for various reactions~\cite{Schweidtmann.2018}, including pharmaceutical product development~\cite{Sano.2020}, electrochemical reductive carboxylation~\cite{Naito.2022}, and polymerization~\cite{Mogilicharla.2015}.
	Based on the surrogate model, a new set of input conditions is proposed for the next experimentation while considering the exploration-exploitation trade-off.
	The goal is to find the input variable values that minimize the objective function.
	TS-EMO extends the Thompson sampling algorithm to the multi-objective optimization setting.
	The promising performance of TS-EMO concerning data efficiency, capacity to handle noise, and the ability for batch-sequential usage~\cite{Bradford.2018} makes the algorithm suitable for the optimization of microgel synthesis.

	As the microgel size is a highly relevant product characteristic in the mentioned applications, we aim to produce microgels of a targeted size (product feature).
	Simultaneously, we optimize the product flow and energy demand (process features) because the synthesis has to meet economic and ecological requirements.
	The synthesis procedure highly influences the characteristics of microgels, and multiple influences on the microgel size have been discovered experimentally.
	The surfactant~\cite{McPhee.1993, Wu.1994, Andersson.2006, Destribats.2014, Wedel.2017, Wolff.2018, Nessen.2013}, monomer~\cite{Virtanen.2014}, cross-linker~\cite{Wu.1994, Balaceanu.2011, Schneider.2014, Virtanen.2019}, and initiator~\cite{Virtanen.2014, Imaz.2008, Chiu.1995} concentration in the synthesis impact the microgel size.
	Also, the process conditions, including reactor temperature~\cite{Wu.1994, Virtanen.2014} and flow profile~\cite{Virtanen.2014, Kather.2018}, determine the microgel size.
	For the synthesis of microgels with constant cross-linking fraction, we include the reaction temperature, initiator and monomer flow, and the surfactant concentration as variable inputs in our data-driven study.

	Since TS-EMO is a stochastic optimization algorithm, it does not guarantee finding the global optimum. 
	We therefore conduct a computational validation step via global deterministic optimization using our open-source software MAiNGO (McCormick-based Algorithm for mixed-integer Nonlinear Global Optimization)~\cite{Bongartz.2018}.
	MAiNGO has been demonstrated to be very suitable for optimization with GPs embedded~\cite{Schweidtmann.2021}.
	The global deterministic optimization ensures that for a given GP and acquisition function the optimal solution is found.
	The computed Pareto-optimal solutions are computed based on the GPs trained on the experimental data.
	Thus, the Pareto-optimal points are estimates and need to be validated experimentally to show that we are truly able to synthesize the desired microgel and to ensure that computational prediction and real experiment agree.
	Therefore, in addition, we validate our optimization results experimentally.
	We conduct the proposed synthesis of a selection of Pareto-optimal points and compare the experimental outcome to the computed findings.

	We structure the remaining manuscript as follows.
	Section~\ref{sec:exp} describes the experimental setup of the microgel synthesis in the flow reactor.
	Section~\ref{sec:opt} reports our optimization approach, including the TS-EMO algorithm, the initial data set, and the problem setup using MAiNGO.
	Section~\ref{sec:res} presents the results of the optimization studies and the computational and experimental validation.
	Finally, we conclude our work in Section~\ref{sec:concl}.

	\section{Experimental}\label{sec:exp}
	
	\subsection{Materials}
	\textit{N}-isopropylacrylamide (NIPAM) (97\%, ITC Chemicals) is distilled under vacuum for purification and recrystallized from hexane. 
	2,2'-azobis(2-methyl\-propion\-amidine)di\-hydro\-chloride (AMPA) (97\%, Sigma-Aldrich), \textit{N},\textit{N}'-methylene\-bis\-(acryl\-amide) (BIS) (99\%, Sigma-Aldrich), and hexa\-decyl\-tri\-methyl\-ammonium bromide (CTAB) ($\geq$97\%, Merck) are used as received. 
	Deionized water (referred to as water) is produced in-house (conductivity \SI{0.8}{\micro\siemens\per\centi\metre} at \SI{25}{\degreeCelsius}).
	
	\subsection{Microgel synthesis in flow reactor}
	We synthesized microgels via precipitation polymerization inside a tubular glass reactor setup, as described in detail in our previous publication~\cite{Kaven.2021}.
	In the following, we provide a brief summary of this experimental setup.
	Two feed solutions are created, where the monomer and initiator are dissolved in water.
	The monomer solution contains deionized water with \SI{110.6}{\milli\mole\per\liter} of NIPAM, \SI{2.7}{\milli\mole\per\liter} of cross-linker BIS, and \SI{0.41}{\milli\mole\per\liter} of surfactant CTAB.
	Thus, the resulting microgels contain a cross-linker fraction of \SI{2.5}{\mole\percent}.
	The initiator solution comprises deionized water with \SI{1.5}{\milli\mol\per\liter} of initiator AMPA.
	Both solutions (initiator and monomer) and constantly degassed using nitrogen.
	The flow rates of the monomer and initiator solution can be controlled between \SIrange{2}{18}{\milli\liter\per\minute} and \SIrange{0.1}{0.9}{\milli\liter\per\minute}, respectively.
	Hence, the overall flow rate and the ratio between both feed flows can be adapted.
	An external heating bath heats the reactor to reaction temperature.
	We adjust the reactor temperature between \SIrange{60}{80}{\degreeCelsius}.
	The produced microgels exit the reactor, and the solution is cooled to stop the reaction.
	During the continuous synthesis, we use Raman spectroscopy to determine the weight fraction of the remaining NIPAM ($w_{NIPAM}$) via in-line measurements.
	Raman spectra are recorded in HoloGRAMS (Kaiser Optical Systems, Ann Arbor, Michigan, USA) with cosmic ray correction using an RXN2 Raman Analyzer (Kaiser Optical Systems) and an acquisition time of \SI{40}{\second}.
	More details on the Raman measurement configuration are described in our previous work~\cite{Kaven.2021}.
	We assess the Raman spectra using an evaluation model based on Indirect Hard Modeling~\cite{Kriesten.2008}, which we previously developed~\cite{Kaven.2021}.
	We published the calibration measurements for the model development for transparency and reproducibility~\cite{data.Kaven.2021}.
	In an off-line step, we use the Zetasizer Ultra (Malvern Panalytical, Malvern, UK) to determine the hydrodynamic diameter ($D_H$) of the collapsed microgels via Dynamic Light Scattering (DLS).
	The microgel samples are diluted in ultrapure water and prepared in a disposable capillary cell of the type DTS0012 for the DLS measurements.
	The measurements are carried out at \SI{50}{\degreeCelsius} with an angle of \SI{90}{\degree} (side scatter).
	Each measurement is repeated four times, and the software ZS Xplorer analyzes the results.
	We exclude experimental data points where the DLS measurements are unreliable due to a high relative error of the microgel size or an increased polydispersity index, indicating that no microgels formed.
	
	\section{Computational}\label{sec:opt}
	
	The following section is structured as follows.
	First, we formulate the optimization problem considering the goals and limitations of the experimental setup, see Sec.~\ref{sec:opt_problem}.
	In Sec.~\ref{sec:initial_data}, we describe the procedure for generating a set of experiments to initialize the iterative optimization study.
	Next, we outline the conducted optimization studies in a high-level description in Sec.~\ref{sec:gen_approach}.
	Further, we give details on the basic operating principle of the employed TS-EMO algorithm in our hardware-in-the-loop setup and the validation approach via global deterministic optimization and the optimization problem definition therein in Sec.~\ref{sec:tsemo} and \ref{sec:maingo}, respectively.

	\subsection{Optimization problem definition}\label{sec:opt_problem}
	The optimization aims to find optimal settings for the synthesis to generate a high product output at short residence times and precise, targeted microgel sizes while minimizing the reaction temperature at steady-state.
	Furthermore, the objectives must be determined from outputs quantifiable via established monitoring techniques.
	
	The reaction system has four optimization variables as inputs $\textbf{x}$: reaction temperature $T$, surfactant concentration $c_{CTAB}$, and flow rates of the initiator $F_I$ and monomer $F_M$ solution.
	The bounds on the inputs are presented in Tab.~\ref{tbl:input_bounds}.
	The range of $T$ comprises the minimum of \SI{60}{\degreeCelsius} when the initiator decomposition effectively sets in~\cite{SigmaAldrich.2023} and the maximum of \SI{80}{\degreeCelsius} when solvent evaporation becomes an issue.
	The bounds on $c_{CTAB}$ are based on the reaction experience that no colloidal stability sets in below the lower limit.
	Generally, a higher $c_{CTAB}$ causes a smaller microgel size.
	Thus, we determined the upper limit for $c_{CTAB}$ based on preliminary experiments.
	The pump's capacity defines the limits for the monomer and initiator solution flow rates.
	Furthermore, at the minimum $F_M = $~\SI{2}{\milli\liter\per\minute}, which entails the maximum residence time in the reactor (approximately~\SI{1800}{\second}), the final conversion is reached, as discovered in our previous work~\cite{Kaven.2021}.
	The employed upper bounds allow for achieving the desired microgel size range, as we conclude from empirical knowledge.
	The concentration of the monomer NIPAM ($c_{NIPAM} =$~\SI{110.6}{\milli\mole\per\liter}) in the stock solution, and the ratio of monomer to cross-linker BIS are kept constant for the reaction optimization to maintain a cross-linking fraction of \SI{2.5}{\mole\%} within the microgel.
	\begin{table}[ht]%
		\caption[Table]{Bounds on input variable values.}
		\label{tbl:input_bounds}
		\centering
		\begin{tabular}{lccc} 
			\hline
			Variable & Unit & Lower bound & Upper bound \\ 
			\hline
			$F_I$ & \SI{}{\milli\liter\per\minute} & 0.1 & 0.9 \\
			$F_M$ & \SI{}{\milli\liter\per\minute} & 2 & 18\\
			$c_{CTAB}$ & \SI{}{\milli\mole\per\liter} & 0.14 & 0.41\\
			$T$ & \SI{}{\degreeCelsius} & 60 & 80 \\
			\hline
		\end{tabular}
	\end{table}
	
	We measure two quantities of the system at the end of the reaction: The weight fraction of the monomer NIPAM $w_{NIPAM}$ and the microgel's hydrodynamic radius $r_H$.
	From the measurements, we derive two quantities $\textbf{y}$ for the surrogate model data set: The product flow ($F_{Product}$) and the squared deviation from the targeted microgel size ($\Delta r_{H}^2$).
	The product flow characterizes the reactor efficiency and is computed via:
		\begin{equation*}
			F_{Product} = \frac{w_{NIPAM,0} - w_{NIPAM,f}}{w_{NIPAM,0}} \cdot (F_I + F_M ) \label{eq:product_flow},
		\end{equation*}
	where $w_{NIPAM,0}$ and $w_{NIPAM,f}$ denote the initial and final NIPAM weight fraction.
	
	The output $\Delta r_{H} ^2$ is calculated as the squared difference between the measured and targeted hydrodynamic radius:
		\begin{equation*}
			\Delta r_{H} ^2= (r_{H,measured} - r_{H,target})^2.
		\end{equation*}
	The targeted microgel size in this contribution is a hydrodynamic radius of \SI{100}{\nano\meter} in the collapsed state at~\SI{50}{\degreeCelsius}, as the size range is relevant in medical applications to pass the human cell boundary.
	Previously, it was found that microgels with a hydrodynamic diameter above \SI{800}{\nano\meter} in the swollen state are unsuitable for cellular uptake~\cite{Switacz.2020}.
	This size corresponds to a hydrodynamic radius of approximately \SI{222}{\nano\meter} at the collapsed state.
	Thus, microgels of a hydrodynamic radius of \SI{100}{\nano\meter} are expected to achieve fast cellular uptake kinetics.

	The efficient microgel production targets a low reaction temperature as heating contributes significantly to energy consumption.
	The reaction temperature $T$ is an input to the reactor system; hence, no additional measurement technology is needed. 
	The difference to the minimum allowable temperature (see Tab.~\ref{tbl:input_bounds}) is defined as another objective function:
		\begin{equation*}
			\Delta T = T - T_{min}.
		\end{equation*}
	Technically, the input temperature can be used as an objective function directly.
	However, we use the temperature difference as the objective to scale the temperature values to a similar magnitude as the flow rates and to underline the generality of the method.
	
	The resulting multi-objective optimization problem is summarized in the following:
		\begin{equation*}
			\begin{alignedat}{2}
				\mpmin{ {\bf x} \in [{\bf x}^L,{\bf x}^U]}  & -F_{Product}, \; \Delta r^2_H, \; \Delta T , 
			\end{alignedat}
		\end{equation*}
	where, ${\bf x}=[F_I, F_M, c_{CTAB}, T]$, and ${\bf x}^L$ and ${\bf x}^U$ denote their corresponding lower and upper bounds as presented in Tab.~\ref{tbl:input_bounds}.

	\subsection{Initial data set}\label{sec:initial_data}
\begin{figure*}[ht]%
		\includegraphics[trim={3.5cm 8cm 8.5cm 5.5cm},clip,width=0.6\textwidth ]{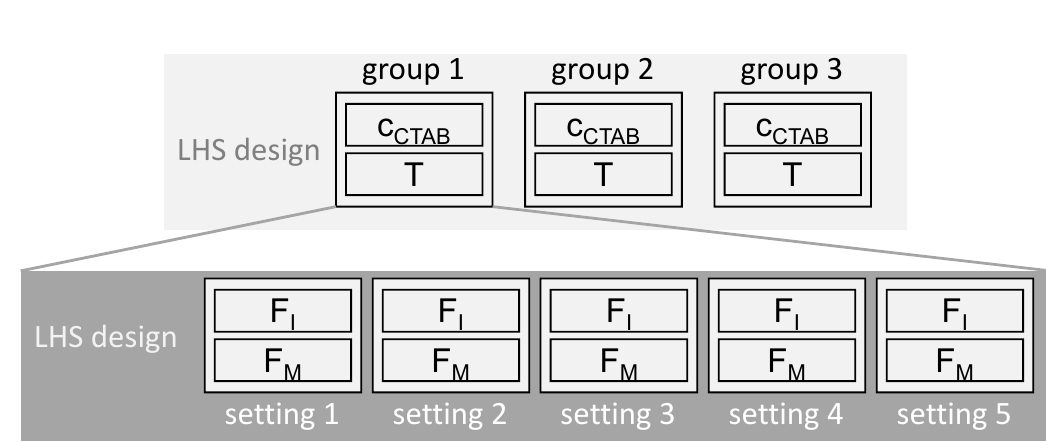}
		\centering
		\caption{Grouping of initial experiments designed via LHS.}%
		\label{fig:lhs}%
	\end{figure*}

	Effective initial values are important to initialize the data-driven optimization algorithm.
	Often random choices are taken as initial guesses, without distinguishing between variables.
	However, we aim for efficient usage of experimental resources and accordingly devised the following tailored initialization.
	We configure three groups of experiments, each comprising five experimental settings.
	The division is visualized in Fig.~\ref{fig:lhs}.
	We distinguish between input variables $T$ and $c_{CTAB}$ that are at a fixed value for each group and inputs $F_M$ and $F_I$ that vary simultaneously within one group.
	We chose five settings per group, as this amount of experimental settings can be conducted within one day of working in the laboratory and is therefore practical.
	Furthermore, we decided to consider three groups of experiments as a trade-off between covering the input space of $T$ and $c_{CTAB}$ sufficiently and conducting a reasonable size of initial experiments in total.
	
	Changing $T$ between experimental runs relates to long transition times.
	Thus, we keep $T$ at a fixed value for each group of experiments for an efficient proceeding.
	Also, $c_{CTAB}$ is fixed for an experimental group, as preparing the monomer solution with different content of CTAB for each experiment execution is laborious and increases the risks of inserting air into the reactor system (oxygen inhibits the reaction) while decreasing the flexibility of the reaction setup.
	Therefore, keeping $c_{CTAB}$ fixed constitutes a trade-off between effort for the synthesis preparation, risk of contamination, and loss of flexibility in synthesis execution.
	
	We employ the \texttt{lhsdesign} function for Latin Hypercube Sampling (LHS) in MATLAB 2019b to determine the input values for the initial experiments.
	In the first step, we set the values for $T$ and $c_{CTAB}$ for each of the three groups via LHS.
	Subsequently, we perform LHS again for $F_I$ and $F_M$ within each group for five settings.
	In total, we derive an amount of 15 initial experiments.

	\subsection{General approach}\label{sec:gen_approach}
	We conduct a hardware-in-the-loop optimization study involving TS-EMO and a validation study including computational and experimental validation.
	In the hardware-in-the-loop approach, we employ TS-EMO to determine the next group of experiments based on an initial experimental data set.
	After the suggested group conditions are experimentally tested, we repeat the optimization process and subsequent experimentation until eleven iterations have been reached.
	Finally, we validate the results from the TS-EMO study computationally via global deterministic optimization using the software MAiNGO and experimentally with reaction settings from Pareto optimal points.

	\subsubsection{TS-EMO algorithm}\label{sec:tsemo}
\begin{figure*}[ht]%
		\includegraphics[trim={1.2cm 1cm 0cm 1.2cm},clip,width= 0.9 \textwidth ]{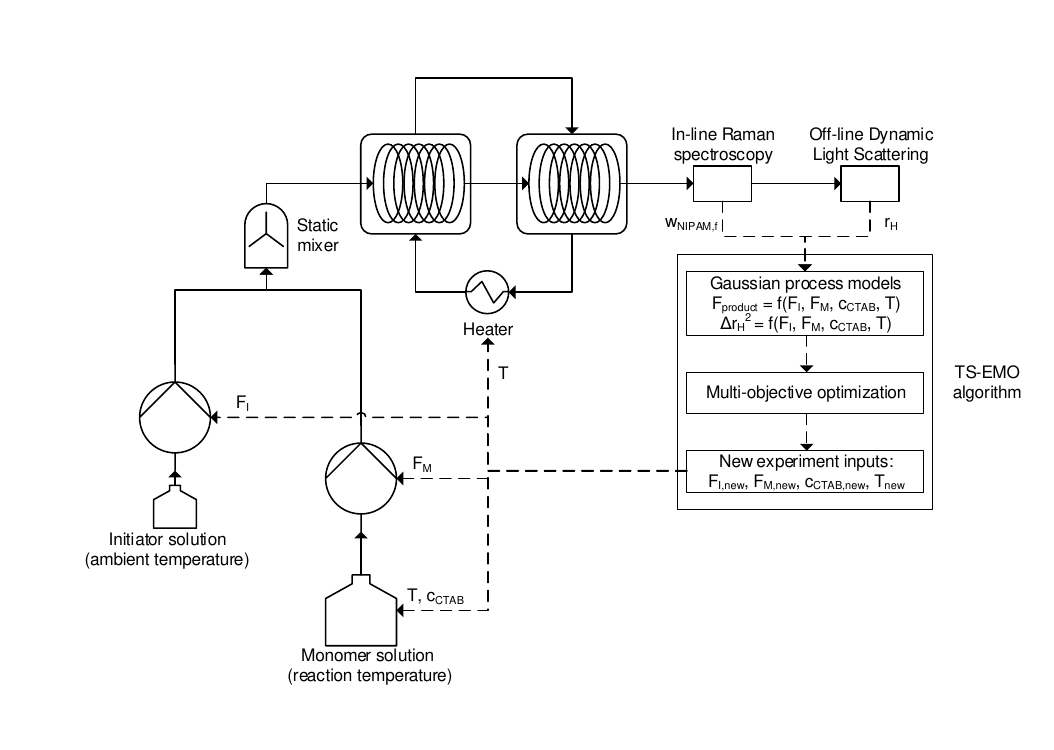}
		\centering
		\caption{Overview of the iterative multi-objective optimization of the microgel synthesis in flow using the TS-EMO algorithm. Solid arrows indicate material flow, dotted arrows represent information transfer.}%
		\label{fig:overview}%
	\end{figure*}

	We apply TS-EMO~\cite{Bradford.2018} to the product-process optimization of the continuous microgel synthesis.
	The schematic setup of the reactor combined with the algorithm is shown in Fig.~\ref{fig:overview}.
	TS-EMO uses experimental data points $\textbf{x}^{(i)} = [F_I, F_M, c_{CTAB}, T]$ and $\textbf{y}^{(i)} := [F_{product}, \Delta r_H^2]$  to create an approximation via a GP surrogate model of the unknown function $f$. 
	For the training of the GPs, we apply matern type 1 as the function kernel.
	The third objective can directly be calculated from the input variables.
	In the multi-objective optimization step, Thompson sampling allows approximating the Pareto set of the optimal solutions.
	Here, we set the number of spectral sampling points to 4,000.
	Lastly, an optimal candidate set of input conditions $\textbf{x}^{(i+1)} = [F_{I,new}, F_{M,new}, c_{CTAB,new}, T_{new}]$ is calculated to continue in the next experimental iteration loop.
	The settings incorporate a genetic algorithm with 1,000 generations for optimization.
	
	In conclusion, the TS-EMO algorithm is provided with an initial experimental data set designed via LHS.
	The algorithm then provides a new set of experiments to be conducted in the following experimental round.
	Subsequently, in each optimization round, we determine a set of the following five experiments at one fixed $T$ and $c_{CTAB}$ with varying $F_I$ and $F_M$.
	We chose batch-sequential optimization, meaning evaluating multiple points in each iteration, as off-line DLS measurements are conducted more efficiently in batch preparation.
	In addition, we chose five experimental settings, as we can adequately conduct this quantity within one day of synthesis experimentation.
	The TS-EMO calculation and the experimentation are repeated in multiple iterations.
	Meanwhile, searching for the optimal recipe should take as few iterations as possible to decrease the experimental effort and expense of chemicals used.
	Thus, the hardware-in-the-loop procedure ends when a certain number of iterations have been performed or the executor decides that sufficient reaction knowledge has been gathered.
	In the presented study, we end the procedure after eleven iterations.

	\subsubsection{Global deterministic optimization}\label{sec:maingo}
	
	For the computational validation, we use MAiNGO~\cite{Bongartz.2018} to conduct a global deterministic optimization where $F_{Product}$ acts as the single objective.
	Additionally, we apply the $\epsilon$-constraint method~\cite{Ehrgott.2009} to restrict the objective $\Delta r_H^2$.
	As the remaining objective $\Delta T$ is directly proportional to the input $T$, we restrict the upper bound of the input variable $T$ step-wise.
	For the global optimization, we use the experimental data received in the TS-EMO study and do not perform further experiments in the form of a hardware-in-the-loop approach.
	We set the starting point and the $\epsilon$ values for the optimization based on the results derived from the hardware-in-the-loop study.
	
	We rewrite the optimization problem to a single-objective formulation in reduced space:
		\begin{equation}
			\begin{alignedat}{2}
				\mpmin{\bf x \in [{\bf x}^L,{\bf x}^U]}  & -F_{Product} \\
				\mpst & \; \Delta r_H^2 \mple \epsilon
			\end{alignedat}
		\end{equation}
	As stated above, the values for $\epsilon$, ${\bf x}^U$ of $T$, and the starting point are derived from the results of the TS-EMO study.
	
	\section{Results and discussion}\label{sec:res}
	
	The results and discussion are organized as follows.
	First, we present our findings from the study involving TS-EMO with four inputs and three objectives in Sec.~\ref{sec:main}.
	There we show the Pareto optimal solutions for the three-dimensional objective system, the progression of the experimental outcome with accumulating experimentation, the error analysis of the measurements, and the Pareto optimal solutions for each of the four inputs.
	Subsequently, we display the results of the validation studies in Sec.~\ref{sec:validation}.
	The computational validation via global deterministic optimization is shown in Sec.~\ref{sec:maingo_res}.
	We re-formulated the optimization problem to a single objective with four input variables for the final study.
	In Sec.~\ref{sec:exp_validation}, we additionally exhibit the experimental validation of three Pareto optimal points.
	We provide all experimental data~\cite{data.Kaven.2023}.
	The data includes the raw Raman measurements and an evaluation of the DLS measurements.
	In addition, we make data points underlying the graphical representation of the results available in Supporting Information Sec.~2.
	The data points include the experimental data (Supporting Information Sec.~2.1) and the Pareto optimal solutions calculated via global deterministic optimization in the validation step (Supporting Information Sec.~2.2).
	As the Pareto optimal solutions calculated via TS-EMO are exhaustive, the data is not provided explicitly.
	The results can be re-constructed by applying TS-EMO on the experimental data provided.
	The software employed in this contribution is available open-source: TS-EMO~\cite{tsemo} and MAiNGO~\cite{maingo} with MeLOn~\cite{melon}, the interface for embedded machine-learning models.

	\subsection{Hardware-in-the-loop involving TS-EMO}\label{sec:main}
	
	We conduct eleven iterations for the hardware-in-the-loop optimization.
	We analyze the Pareto optimal solutions in detail regarding the feasible space of the objective values in Sec.~\ref{sec:pareto}.
	Next, the progression of the experimentation outcome, an analysis of the errors from the experimental measurements, and the computational uncertainty of the calculated Pareto front are presented in Sec.~\ref{sec:progression}.
	Lastly, we evaluate the input variable values at the Pareto optimal points to derive suitable reactor settings for the desired microgel product in Sec.~\ref{sec:pareto_inputs}.

	\subsubsection{Pareto optimal solutions}\label{sec:pareto}
\begin{figure}[ht]	
		\includegraphics[width=0.6\textwidth ]{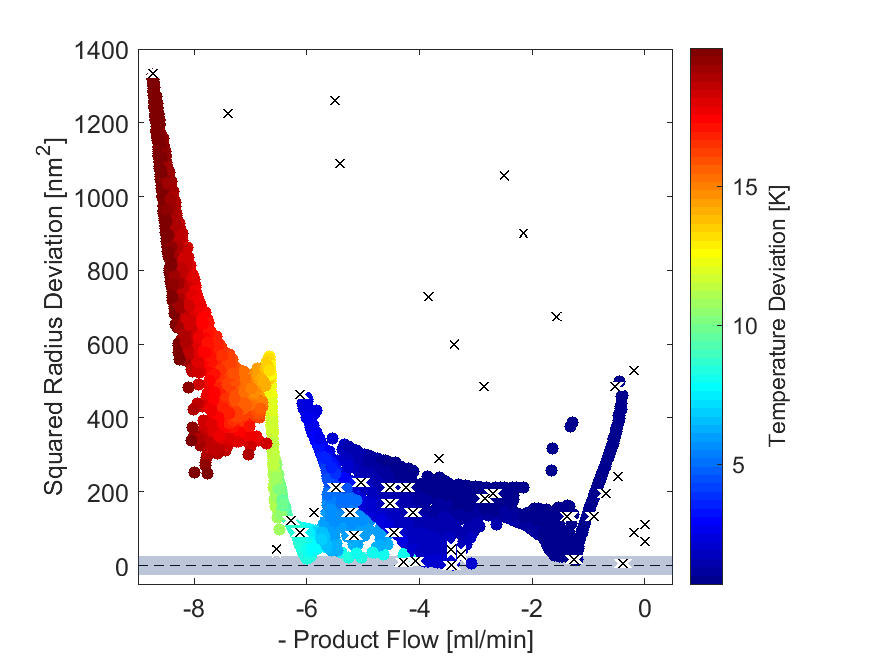}
		\centering
		\caption{Estimated Pareto front of the hardware-in-the-loop study using TS-EMO: Squared radius deviation over product flow.
			The color scale indicates the temperature deviation.
			The x symbols mark the experimental data points.
			The shaded area maps the deviation of $\pm \SI{5}{\nano\meter}$ to the desired microgel radius.}
			\label{fig:SizeOverFlow_TempDev}
	\end{figure}

	In the hardware-in-the-loop study, $F_I$, $F_M$, $c_{CTAB}$, and $T$ are varied as the inputs to the reactor, and $F_{Product}$, $\Delta r_H^2$, and $\Delta T$ are the objectives.
	Fig.~\ref{fig:SizeOverFlow_TempDev} shows the resulting Pareto front of the study.
	We used a population size of 5,000 to represent the three-dimensional Pareto front sufficiently.
	As visualizing three objectives is challenging, we proceed with a two-dimensional plot and add a color scale for the third objective to visualize the estimated Pareto front for better interpretation.
	However, it is crucial to remember that we are considering three-dimensional optimization results for the meaningful interpretation of the two-dimensional plots.
	
	For the two-dimensional Pareto fronts, the desired outcome in Fig.~\ref{fig:SizeOverFlow_TempDev}, the utopia point, of the multi-objective optimization regarding the product flow and the squared radius deviation is located in the bottom left corner of the plot.
	Equally, small temperature deviations (depicted in dark blue) indicate the location of the utopia point in the third dimension.
	Looking at the results, it appears that the three objective functions are conflicting; thus, reaching the utopia point is impossible. 
	In other words: the product flow rate becomes lower for microgels closer to the targeted size, and higher temperatures are needed for high product flow rates.
	In addition, the shaded area around a squared radius deviation accounts for a difference of $\pm \SI{5}{\nano\meter}$ or $5\%$ to the desired size.

	The analysis of the estimated Pareto front in Fig.~\ref{fig:SizeOverFlow_TempDev} yields that up to \SI{6.0}{\milli\liter\per\minute} of product flow, a microgel size sufficiently close ($\pm \SI{5}{\nano\meter}$) to the desired size is achievable.
	The microgel size deviation begins to diverge more strongly from the targeted value after a product flow rate of approximately \SI{6.5}{\milli\liter\per\minute} is reached.
	This deviation shows that product flow rates above a value of around \SI{6.5}{\milli\liter\per\minute} are incompatible with the targeted microgel size.

	The temperature influences the optimal product flow more significantly than the optimal microgel size.
	This trend is represented by the color indicated temperature change that is more substantial along the x-axis than the y-axis.
	The underlying GPs (depicted in Supplementary Information Sec.~1) show that an increase in temperature generally accompanies an increase in product flow.
	Still, the product flow converges towards approximately \SI{6.5}{\milli\liter\per\minute} for temperatures above approximately \SI{70}{\degreeCelsius} (corresponding to \SI{10}{\kelvin} temperature deviation).
	Thus, low temperatures (below~\SI{70}{\degreeCelsius}) are sufficient considering the trade-off between maximizing product flow and achieving the targeted microgel size, as above approximately \SI{70}{\degreeCelsius} only the product flow improves.
	Overall, the optimal temperature input spans the entire allowable range between \SIrange[]{60}{80}{\degreeCelsius}.
	Furthermore, the GP for the squared radius deviation (Supplementary Information Sec.~1) shows an increase with rising temperatures.
	However, the correlation between reaction temperature and microgel size deviation appears highly non-linear and subject to inherent variance.
	Lastly, the underlying GP for the temperature deviation (Supplementary Information Sec.~1) confirms the successful training of the GPs, as the temperature deviation shows no correlation to $F_I$, $F_M$, or $c_{CTAB}$, and is directly proportional to the input temperature values with little variance.

	In conclusion, the results concerning a suitable microgel size at a high product flow and medium reactor temperatures are promising.
	The underlying GPs confirm our apriori reaction knowledge; thus, we can validate the functionality of the applied method elementarily.
	However, the GPs are occasionally subject to high variance, and the available data points are limited.
	Nevertheless, we can derive meaningful information about the synthesis, e.g., limiting the temperature to \SI{70}{\degreeCelsius} is sufficient for successful synthesis.
	Furthermore, we find that a maximum product flow of \SI{6.0}{\milli\liter\per\minute} is achievable when restricting the allowable microgel size deviation to $\pm \SI{5}{\nano\meter}$.

	\subsubsection{Experiment progression and error analysis}\label{sec:progression}
	
\begin{figure}[ht]	
		\includegraphics[width=0.8\textwidth ]{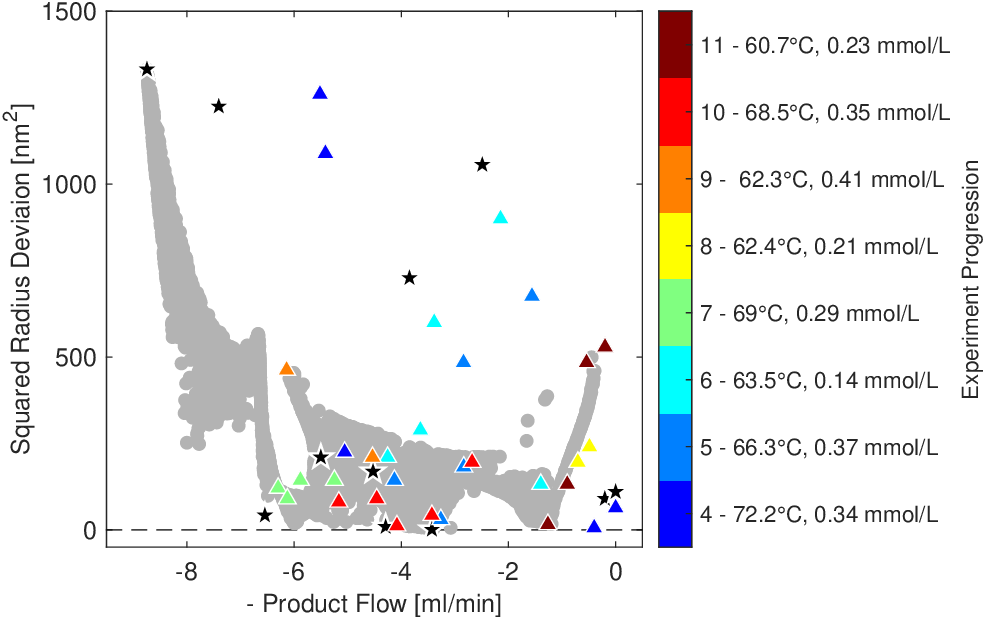}
		\centering
		\caption{Estimated Pareto front of the hardware-in-the-loop study using TS-EMO: Squared radius deviation over product flow. 
			The gray circles represent the estimated Pareto optimal solutions based on the GPs.
			The stars indicate the initial experimental data set and the triangles the subsequent experimental data points, while the color of the triangles shows the experimental progression.}%
		\label{fig:pareto_progression}
	\end{figure}

	In Fig.~\ref{fig:pareto_progression}, the calculated Pareto front is shown with the progression of the experimentation.
	The temperature and the surfactant concentration for each experimental group are listed in addition to the order of experiment progression on the color scale.
	In the graph, the stars mark the results from the initial experiments designed via LHS.
	The LHS ensures a good distribution over the input space.
	The initial experimental results also cover the output space adequately, indicating that the initial data set already provides a reasonable basis for information on the reaction.
	
	Furthermore, the triangles depicted in a color scale represent the experimentally determined data points and their progression in the hardware-in-the-loop approach.
	In each set of experiments, five data points are received.
	We must neglect some data points due to DLS measurement showing a high size distribution index (indicating that no real microgel was formed) or a high relative measurement error.
	Thus, a reduced amount of experimental data points is shown.
	There is no clear trend visible in the experiment progression, as the algorithm tries to balance exploitation and exploration in the design of the next experiment.
	The listed temperature and surfactant concentration values along the experimental progression show that the algorithm mostly explores temperature regions below \SI{70}{\degreeCelsius}.
	While the surfactant concentration is varied over the entire allowed input space.
	Also, for the conducted experiments in this study, the algorithm does not repeat in any iteration the suggested experimental conditions regarding the combination of temperature and surfactant concentration.
	Although output measurements are sometimes excluded without further information to the algorithm, the algorithm does not try to re-evaluate the correlated input space.
	The batch-sequential procedure presumably achieves that the algorithm carries on without going back to previously tested conditions where no information was received.
	In other words: although no input information is gathered at certain input conditions within one experimental group, the information from the remaining input conditions within the group supports the algorithm enough.

\begin{figure}[ht]	
		\includegraphics[width=0.6\textwidth ]{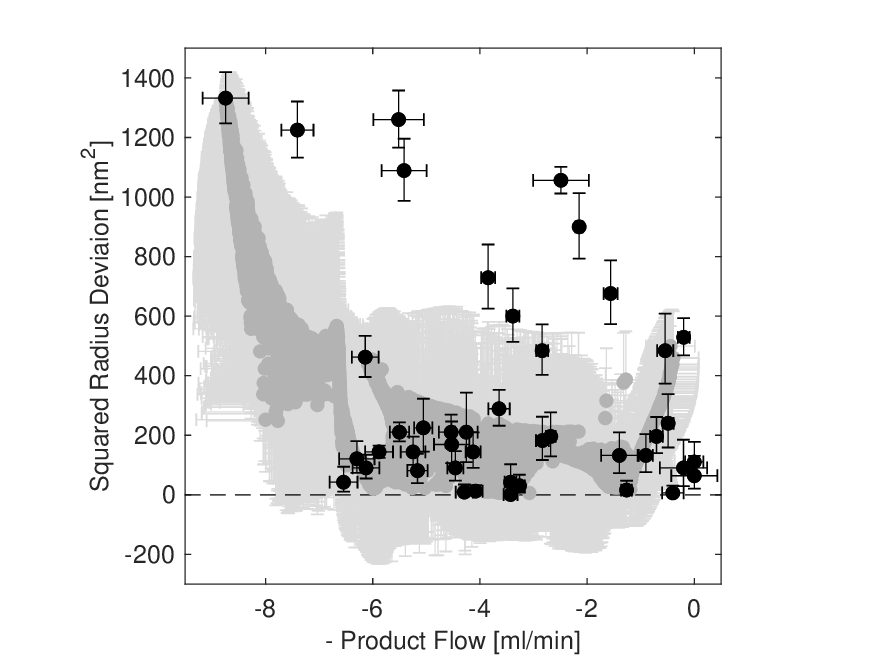}
		\centering
		\caption{Estimated Pareto front of the hardware-in-the-loop study using TS-EMO: Squared radius deviation over product flow. 
			The gray circles represent the estimated Pareto optimal solutions based on the GPs and the according standard deviation.
			The black circles indicate the experimental outcomes and the according measurement uncertainty.}%
		\label{fig:pareto_error}
	\end{figure}

	In Fig.~\ref{fig:pareto_error}, the calculated Pareto front is shown with the computational standard deviation of the optimal points.
	Also, the experimental data points are depicted with the according experimental error bars.
	The magnitude of the experimental error is derived from the measurement technology.
	The evaluation model of the Raman measurements has an inherent root mean squared error of cross-validation (RMSECV) of 0.037~wt-\%.  
	The error propagation, including the RMSECV, is considered for the uncertainty of the product flow.
	For the DLS measurement, the Zetasizer Ultra internally evaluates the standard deviation over the four conducted measurements.
	This error value is also propagated for the uncertainty of the experimental squared particle size deviation.

	Some experimental data points lie slightly below the estimated Pareto front.
	This phenomenon becomes visible in a three-dimensional analysis.
	However, the considered experimental data points lie within the calculated standard deviation of the estimated Pareto front for the squared radius deviation.
	Furthermore, the experimental error bars resulting from the DLS and Raman measurement errors are displayed to underline the magnitude of uncertainty inherent in the real-life experimental setup.

	\subsubsection{Pareto optimal solutions for different inputs}\label{sec:pareto_inputs}
\begin{figure}[!ht]
		\centering
		\subfigure[CTAB concentration]
		{
			\label{subfig:4inputs3obj_CTAB}
			\includegraphics[width=0.45\textwidth]{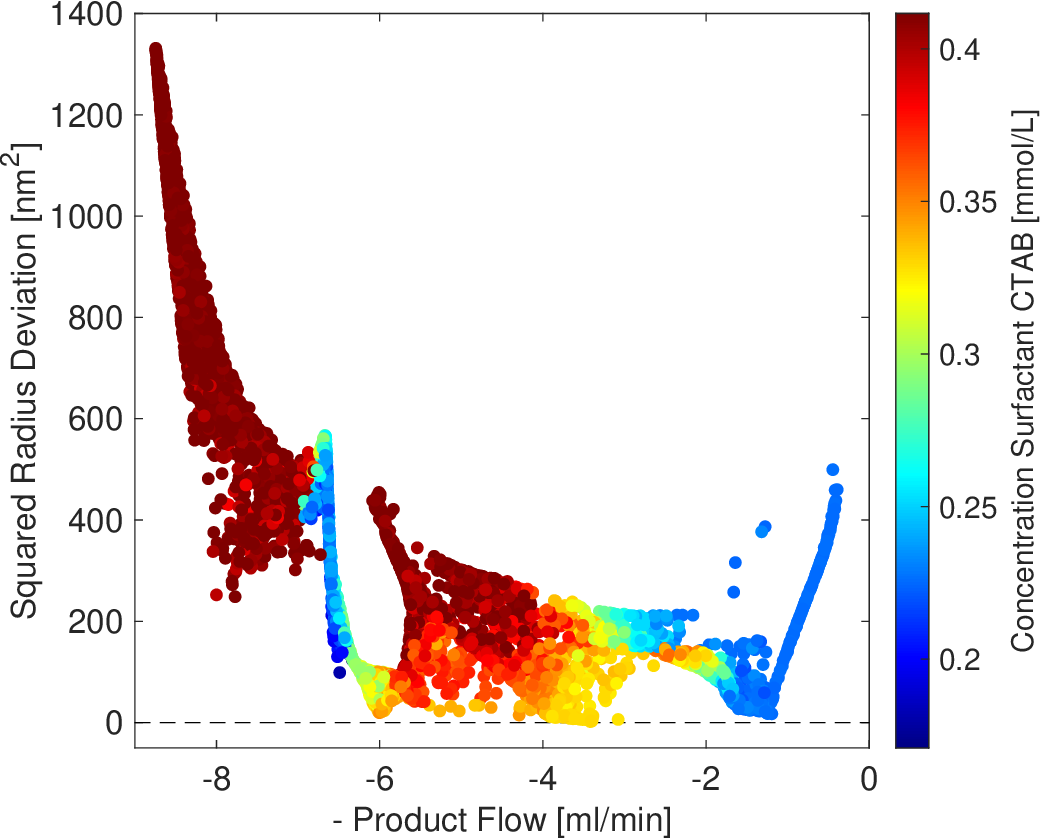}
		}
		\subfigure[Monomer flow rate]
		{
			\label{subfig:4inputs3obj_monomer}
			\includegraphics[width=0.45\textwidth]{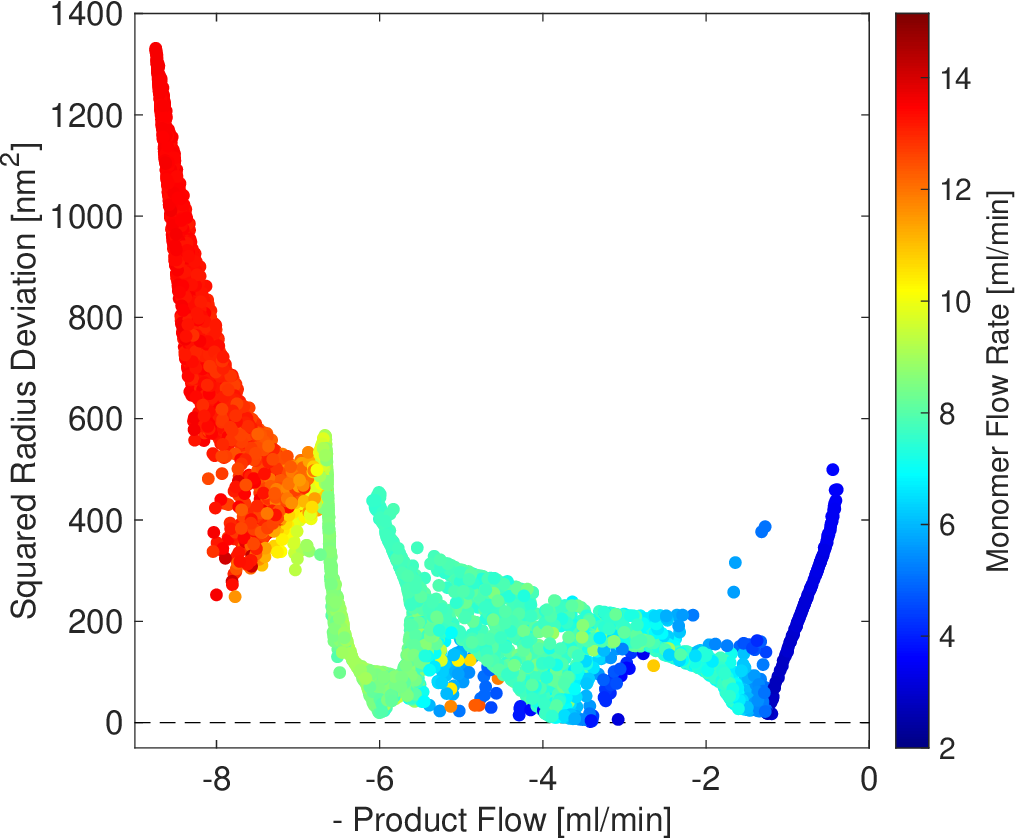}
		}\\
		\subfigure[Initiator flow rate]
		{
			\includegraphics[width=0.45\textwidth]{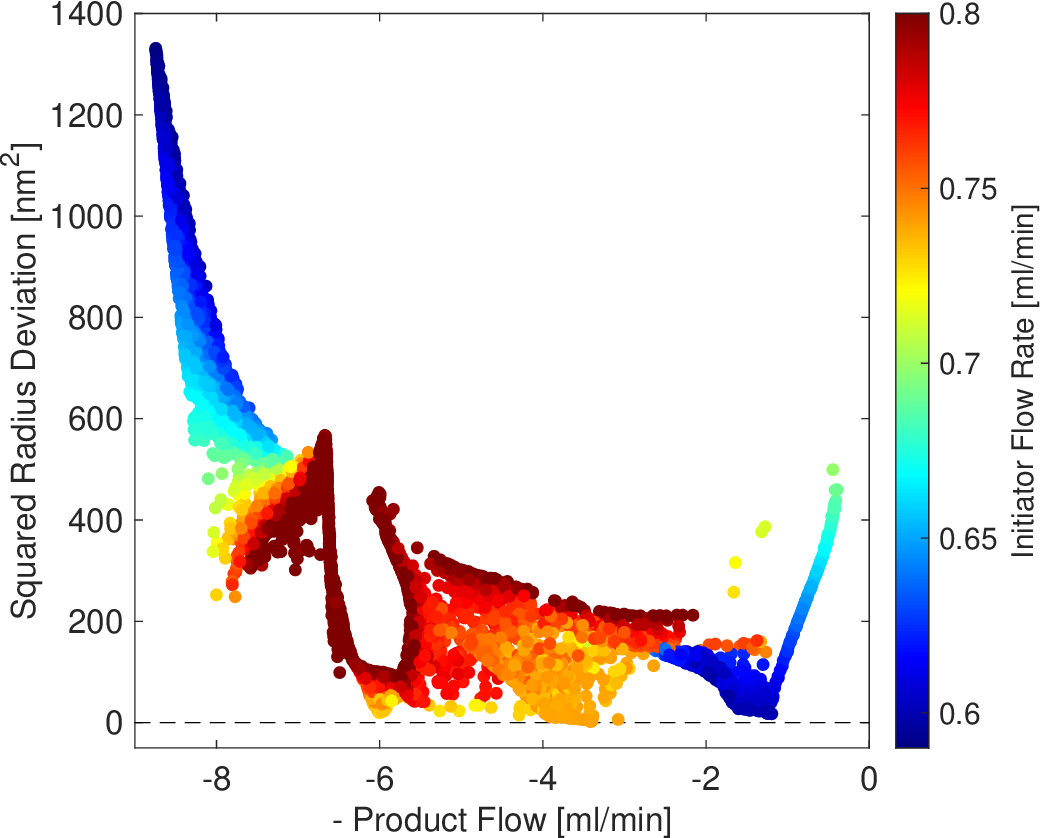}
			\label{subfig:4inputs3obj_initiator}
		}
		\caption{ Estimated Pareto front of main study: Squared radius deviation over product flow for the input variables (a) CTAB concentration, (b) monomer flow rate, and (c) initiator flow rate.
			The circles represent the estimated Pareto optimal solutions based on the GPs, while the color scale indicates the magnitude of the respective input variable.}
		\label{fig:4inputs3obj_inputs}%
	\end{figure}

	In Fig.~\ref{subfig:4inputs3obj_CTAB} to \ref{subfig:4inputs3obj_initiator}, the Pareto front for the objectives $\Delta r_H^2$ and $F_{Product}$ and three out of the four applied inputs is shown.
	The color scale indicates the associated input configuration.
	The inputs pictured include the surfactant concentration, the monomer, and the initiator flow rate.
	The Pareto front with the according input temperature is not depicted explicitly, as Fig.~\ref{fig:SizeOverFlow_TempDev} contains information on the input temperature.
	
	Fig.~\ref{subfig:4inputs3obj_CTAB} shows that the microgel size deviates strongly from the desired size for higher $c_{CTAB}$ values.
	Overall, $c_{CTAB}$ ranges only between \SIrange[]{0.22}{0.41}{\milli\mole\per\liter}.
	The underlying GP (depicted in Supplementary Information Sec.~1) indicates that the product flow can be considered independent of $c_{CTAB}$.
	In contrast, the correlation between squared radius deviation and $c_{CTAB}$ is impaired by high variance.
	The finding that the product flow is unaffected by $c_{CTAB}$ follows the expected outcome, as a change in stabilizer should not impact the conversion kinetics of the reaction system.

	In Fig.~\ref{subfig:4inputs3obj_monomer}, the monomer flow rate ranges between \SIrange[]{2.75}{14.2}{\milli\liter\per\minute} and mainly correlates to the product flow.
	The relation between monomer flow rate and product flow is defined in Eq.~\eqref{eq:product_flow} stating that generally, the monomer flow and product flow are directly proportional (second term in Eq.~\eqref{eq:product_flow}).
	However, the monomer flow rate is also related to the conversion (first term of Eq.~\eqref{eq:product_flow}).
	A higher monomer flow can cause a smaller conversion, as not all monomer can be consumed in the smaller residence time.
	The underlying GP (depicted in Supplementary Information Sec.~1) shows the trade-off between high monomer flow rates associated with an increased overall flow and a lower conversion and low monomer flow rates with a low overall flow but higher conversion.
	Furthermore, the monomer flow rate has little to no influence on the microgel size deviation according to the underlying GP.
	
	Finally, Fig.~\ref{subfig:4inputs3obj_initiator} shows the Pareto front for different initiator flow rates.
	Here, the initiator flow rate ranges between \SIrange[]{0.59}{0.8}{\milli\liter\per\minute} with a clear tendency to the upper bound.
	Similar to the monomer flow rate, the initiator flow rate is directly proportional to the product flow as defined in Eq.~\eqref{eq:product_flow}.
	However, the initiator flow is a maximum of a third of the total flow rate and thus less significant for the overall change in residence time.
	As expected, the underlying GP (depicted in Supplementary Information Sec.~1) also shows a highly linear correlation between initiator flow rate and product flow.
	In addition, the GP for the squared radius deviation shows no clear trend depending on the initiator flow rate.

	\subsection{Validation} \label{sec:validation}
	
	The validation conducted within this contribution includes a computational and experimental part.
	The computational validation is global deterministic optimization of the final GP, Sec.~\ref{sec:maingo_res}.
	The experimental validation is carried out for three calculated Pareto optimal solutions, and the results are shown in Sec.~\ref{sec:exp_validation}.
	
	\subsubsection{Computational validation via global deterministic optimization}\label{sec:maingo_res}
	We proceed with a final deterministic global optimization using MAiNGO.
	The results from the hardware-in-the-loop study are incorporated into the final optimization for validation.
	First, the data points from the TS-EMO study are used to train GPs for $F_{Product}$ and $\Delta r_H ^2$.
	The training settings are the same as for the GPs used in the hardware-in-the-loop approach including TS-EMO.
	Second, the identified optimal point close to the targeted microgel size and a sufficient product flow at a reasonably low temperature is embedded as the starting point of the optimization: $F_I =$ \SI{0.73}{\milli\liter\per\minute}, $F_M =$ \SI{8.1}{\milli\liter\per\minute}, $c_{CTAB} =$ \SI{0.34}{\milli\mole\per\liter}, and $T =$ \SI{68.5}{\degreeCelsius}.
	The calculated outcome for these input variables yields a microgel size deviation of \SI{21.1}{\nm\squared} and a product flow of \SI{6.0}{\milli\liter\per\minute}.
	Also, the visualization of the TS-EMO study (see Fig.~\ref{fig:SizeOverFlow_TempDev}) allows setting reasonable values for the $\epsilon$ constraint method.
	
\begin{figure}[!ht]
		\centering
		\subfigure[CTAB concentration]
		{
			\label{subfig:maingo_CTAB}
			\includegraphics[width=0.45\textwidth]{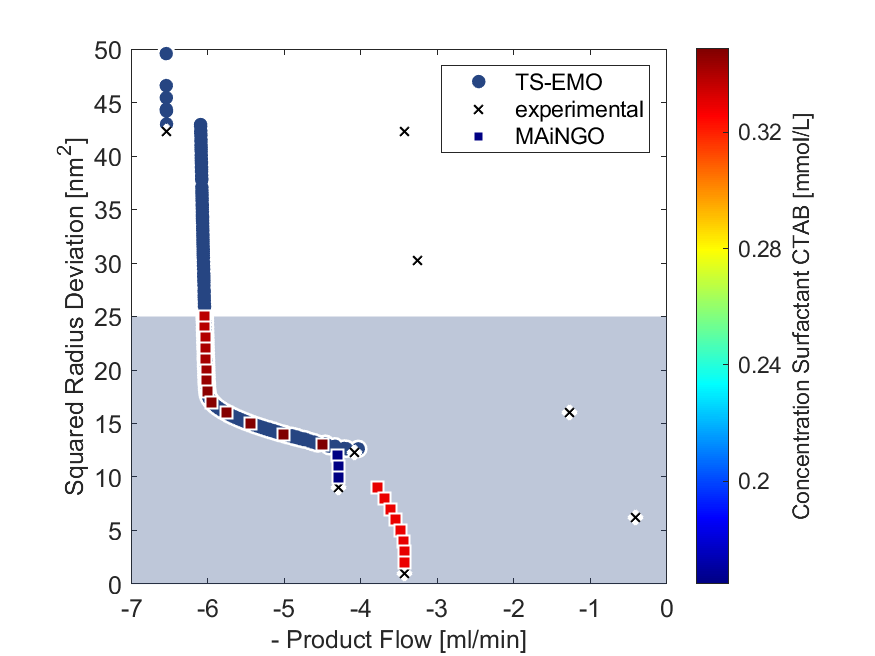}
		}
		\subfigure[Monomer flow rate]
		{
			\label{subfig:maingo_monomer}
			\includegraphics[width=0.45\textwidth]{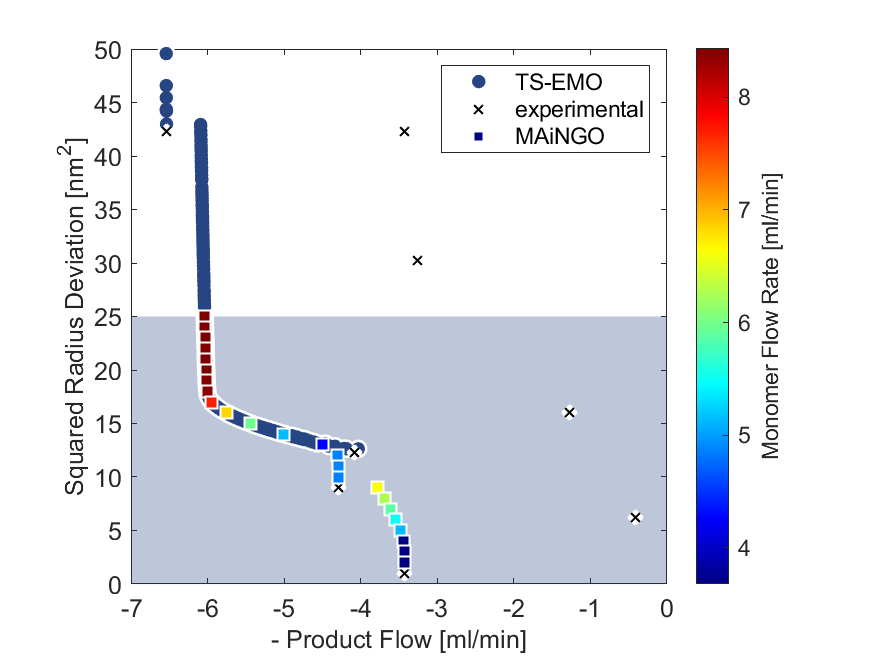}
		}\\
		\subfigure[Initiator flow rate]
		{
			\includegraphics[width=0.45\textwidth]{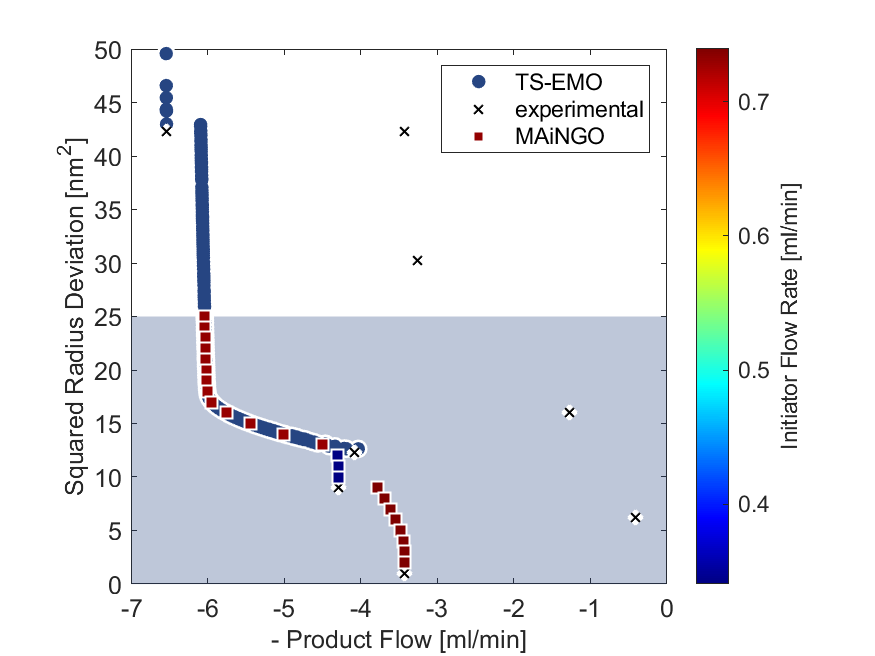}
			\label{subfig:maingo_initiator}
		}
		\subfigure[Reaction temperature]
		{
			\includegraphics[width=0.45\textwidth]{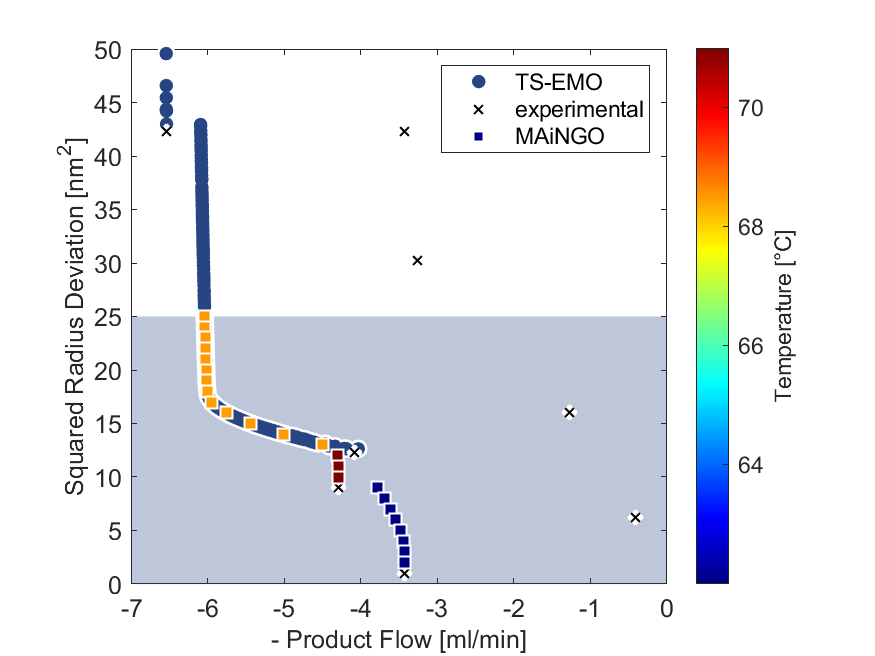}
			\label{subfig:maingo_temp}
		}
		\caption{ Estimated Pareto front of global deterministic optimization: squared radius deviation over product flow for the input variables (a) CTAB concentration, (b) monomer flow rate, (c) initiator flow rate and, (d) reaction temperature.
			The squares represent the estimated Pareto optimal solutions based on the GPs, while the color scale indicates the magnitude of the respective input variable.
			The x symbols mark the experimental data points.
			The blue circles indicate the estimated Pareto front via TS-EMO for two objectives only.}
		\label{fig:maingo_inputs}%
	\end{figure}

	For the deterministic global optimization, the results including the $\epsilon$ constraint method, are presented in Fig.~\ref{fig:maingo_inputs} for each input separately.
	Here, we constrain the squared radius deviation step-wise with a maximum of \SI{25}{\nano\meter\squared}.
	The problem becomes infeasible for squared radius deviations below \SI{2}{\nano\meter\squared}.
	We compare the global deterministic optimization (MAiNGO) with the optimization results for two objectives (product flow and squared radius deviation) using TS-EMO.
	
	Overall, the Figs.~\ref{subfig:maingo_CTAB} to~\ref{subfig:maingo_temp} show that the experimental data points, the Pareto front generated via TS-EMO, and the Pareto front obtained from MAiNGO agree correctly above a product flow of approximately \SI{4.3}{\milli\liter\per\minute}.
	TS-EMO finds a feasible Pareto optimal solution only down to \SI{12.6}{\nano\meter\squared} at a product flow of \SI{4.0}{\milli\liter\per\minute}.
	In this region, the calculated solution via MAiNGO diverges and includes feasible solutions in the product flow range around \SI{4.3}{\milli\liter\per\minute} with squared radius deviations between \SIrange{10}{12}{\nano\meter\squared}.
	
	Within the Pareto optimal solutions calculated via MAiNGO, three regimes can be differentiated most visible for the CTAB concentration and the reaction temperature.
	These regimes range at a product flow of \SIrange[]{3.4}{3.8}{\milli\liter\per\minute}, around \SI{4.3}{\milli\liter\per\minute}, and \SIrange[]{4.5}{6}{\milli\liter\per\minute}.
	In each regime, the CTAB concentration, the initiator flow rate, and the reaction temperature are approximately constant, and only the monomer flow rate varies.

\begin{figure}[ht]	
		\includegraphics[width=0.6\textwidth ]{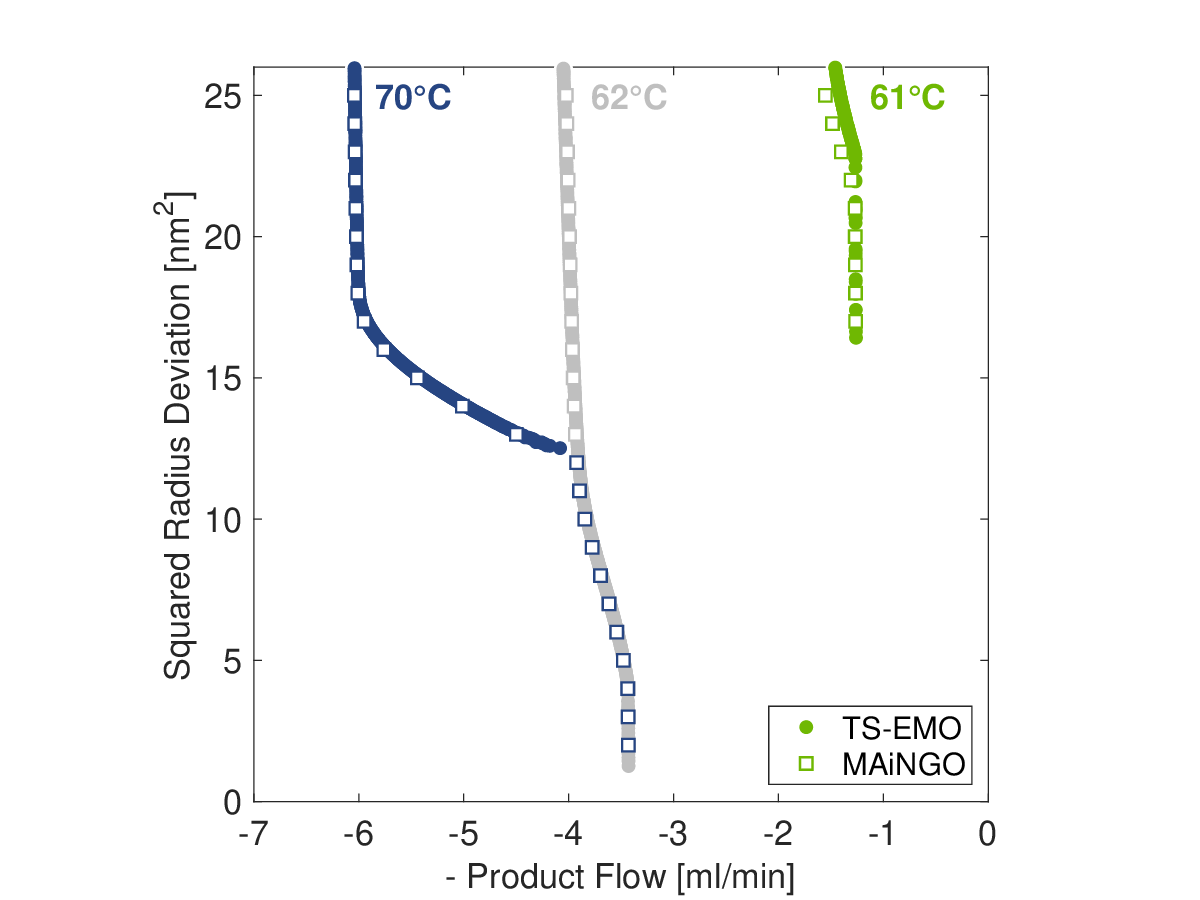}
		\centering
		\caption{Comparison of TS-EMO and MAiNGO results for different bounds on input temperature.
			The filled circles represent the Pareto optimal points calculated via TS-EMO, while the squares show the Pareto optimal points calculated via MAiNGO.}%
		\label{fig:tsemo_maingo_temps}
	\end{figure}

	Further, we change the upper bound of the reactor temperature input variable value to \SI{61}{\degreeCelsius}, \SI{62}{\degreeCelsius}, and \SI{70}{\degreeCelsius}.
	The results of the TS-EMO optimization with two objectives compared to global deterministic optimization results via MAiNGO are shown in Fig.~\ref{fig:tsemo_maingo_temps}.
	The problem becomes infeasible for squared radius deviations below \SI{2}{\nano\meter\squared} for temperatures \SI{62}{\degreeCelsius} and higher, and below \SI{16}{\nano\meter\squared} for \SI{61}{\degreeCelsius}.
	
	In Fig.~\ref{fig:tsemo_maingo_temps}, the Pareto optimal points generated via TS-EMO and MAiNGO agree mostly.
	Only for a maximum input temperature of \SI{61}{\degreeCelsius} the global deterministic optimization via MAiNGO finds slightly better Pareto optimal points for squared radius deviations above \SI{23}{\nano\meter\squared}.
	However, the product flow range between \SIrange[]{1.3}{1.6}{\milli\liter\per\minute} and a minimum squared radius deviation of \SI{16.4}{\nano\meter\squared} for the associated temperature are undesirable.
	Thus, temperatures above \SI{61}{\degreeCelsius} are more relevant.
	For a maximum input temperature of \SI{62}{\degreeCelsius}, the Pareto optimal product flow is limited to \SI{4}{\milli\liter\per\minute} even for substantial deviations in squared radius at \SI{25}{\nano\meter\squared}.
	The Pareto optimal points for squared radius deviations below \SI{13}{\nano\meter\squared} overlap for the MAiNGO and TS-EMO optimization for \SI{62}{\degreeCelsius} and \SI{70}{\degreeCelsius}.
	For a maximum input temperature of \SI{70}{\degreeCelsius}, a notable improvement of the product flow up to approximately \SI{6}{\milli\liter\per\minute} is achievable when allowing squared radius deviations starting at \SI{18}{\nano\meter\squared} and above.
	The TS-EMO Pareto optimal points only cover squared radius deviations above \SI{12.5}{\nano\meter\squared} for a maximum temperature of \SI{70}{\degreeCelsius}.
	The Pareto optimal points for the MAiNGO optimization with a maximum temperature of \SI{70}{\degreeCelsius} and \SI{80}{\degreeCelsius} agree except for the regime around \SI{4.3}{\milli\liter\per\minute} and squared radius deviations of \SIrange[]{10}{12}{\nano\meter\squared}
	indicating that temperatures above \SI{70}{\degreeCelsius} are irrelevant for optimized reactor settings.

	Overall, the Pareto optimal solutions of TS-EMO and MAiNGO agree very well.
	Hence, the hardware-in-the-loop procedure using TS-EMO is validated sufficiently.
	However, the global deterministic optimization finds feasible Pareto optimal solutions beyond TS-EMO.
	The global deterministic optimization of the multi-objective synthesis problem is beneficial because little data is available, and thus guaranteeing a reliable and reproducible solution is crucial.
	However, the surrogate models represented by GPs are subject to significant variance.
	Thus, a solution representing the actual reality remains challenging.
	We also demonstrate that the deterministic single-objective formulation is advantageous here to focus on the output space of interest and reduce computational effort.

	\subsubsection{Experimental validation} \label{sec:exp_validation}
	We conduct three experiments along the deterministically estimated Pareto front for an experimental validation step to determine if the computed estimated based on the trained GPs can be verified experimentally. 
	The inputs, the estimated, and experimentally determined values are presented in Tab.~\ref{tbl:exp_validation}.
	The experimental and calculated values agree very well for the product flow.
	The most significant difference regarding the product flow occurs in Experiment 3 with an absolute divergence of \SI{0.03}{\milli\liter\per\minute} (or approximately 2.8\%) to the calculated value.
	Generally, the agreement of calculated and experimental values is higher for the product flow than for the squared radius deviation.
	The most notable difference regarding the squared radius deviation arises for Experiment 1, where the absolute divergence is \SI{83}{\nano\meter\squared}.
	This significant divergence can be attributed to the high variation in the GP prediction for the squared radius deviation.
	At the same time, the estimated and experimental value for Experiments 2 and 3 agree sufficiently.
	Experiment 3 shows that we can efficiently synthesize microgels with a radius of \SI{101.5}{\nano\meter}, which is acceptable in terms of accuracy.
	
	Overall, the experimental validation indicates that the obtained data is enough to enable an adequate prediction via a GP surrogate model.
	The agreement between estimated and calculated data is good, although the underlying GPs are subject to significant variance.
	The applied procedure is successful with an absolute deviation of \SI{1.5}{\nano\meter} to the desired microgel radius.
	
	\begin{table*}[ht]%
		\caption[Table]{Experimental validation of global deterministic optimization.}
		\label{tbl:exp_validation}
		\centering
		\begin{tabular}{cccccc} 
			\hline
			\multirow{2}{*}{Experiment} & \multirow{2}{*}{Input} & \multirow{2}{*}{Value} & \multirow{2}{*}{Output} & Estimated & Experimental \\ 
			& & & & value & value \\
			\hline
			\multirow{4}{*}{1}    & $T$           & \SI{68.5}{\degreeCelsius}             &                 & & \\
			& $c_{CTAB}$    & \SI{0.35}{\milli\mole\per\liter}      & $F_{Product}$   & \SI{5.95}{\milli\liter\per\minute} & \SI{5.93}{\milli\liter\per\minute} \\
			& $F_I$         & \SI{0.73}{\milli\liter\per\minute}    & $\Delta r_H ^2$ & \SI{17}{\nano\meter\squared} & \SI{100}{\nano\meter\squared} \\
			& $F_M$         & \SI{7.69}{\milli\liter\per\minute}    &                 & &\\
			\hline
			\multirow{4}{*}{2}    & $T$           & \SI{71.0}{\degreeCelsius}             &                 & & \\
			& $c_{CTAB}$    & \SI{0.16}{\milli\mole\per\liter}      & $F_{Product}$   & \SI{4.29}{\milli\liter\per\minute} & \SI{4.20}{\milli\liter\per\minute} \\
			& $F_I$         & \SI{0.34}{\milli\liter\per\minute}    & $\Delta r_H ^2$ & \SI{10}{\nano\meter\squared} & \SI{12.25}{\nano\meter\squared} \\
			& $F_M$         & \SI{4.87}{\milli\liter\per\minute}    &                 & &\\
			\hline
			\multirow{4}{*}{3}    & $T$           & \SI{62.0}{\degreeCelsius}             &                 & & \\
			& $c_{CTAB}$    & \SI{0.33}{\milli\mole\per\liter}      & $F_{Product}$   & \SI{3.43}{\milli\liter\per\minute} & \SI{3.53}{\milli\liter\per\minute} \\
			& $F_I$         & \SI{0.74}{\milli\liter\per\minute}    & $\Delta r_H ^2$ & \SI{2}{\nano\meter\squared} & \SI{2.25}{\nano\meter\squared} \\
			& $F_M$         & \SI{3.68}{\milli\liter\per\minute}    &                 & &\\
			\hline
		\end{tabular}
	\end{table*}
	
	\section{Conclusions}\label{sec:concl}
	Polymerization reactions in flow reactors play an essential role in precise polymer production.
	The efficient, accurate, reproducible synthesis of polymers such as microgels is important.
	Data-driven optimization supports the microgel development effectively.
	We incorporate the multi-objective optimization algorithm TS-EMO to optimize the synthesis of tailored microgels ecologically and economically.
	The proposed synthesis settings enable a product flow of maximum \SI{6.0}{\milli\liter\per\minute} while remaining in an acceptable range of $\pm \SI{5}{\nano\meter}$ to the targeted hydrodynamic radius.
	We use the global deterministic optimization software MAiNGO to prove the reliability and reproducibility of the results.
	In addition, we demonstrate the usefulness of global deterministic solutions for problems with little data availability.
	
	From the experimental side, including Raman spectroscopy constitutes a powerful in-line process analytical tool that has the potential to be incorporated into automated reaction optimization setups.
	Limitations of the proposed work include the non-automated reactor system due to off-line DLS measurements.
	Dependable in-line size determination remains a critical shortcoming on the road to autonomous reaction optimization.
	Furthermore, the DLS data is occasionally unreliable or shows a high polydispersity (indicating no real microgel is produced).
	At the moment, these data points are discarded but could be meaningfully included as valuable information for the algorithm in the future.
	The reliability of DLS data and the challenging interpretation of the GP predictions shows that expert knowledge is still crucial in the optimization procedure and limits a potentially autonomous process based on machine learning.
	Generally, data-driven optimization is limited to a specific reactor setup.
	However, we can quickly adapt the proposed framework to other desired microgel properties and reactor setups.
	Thus, this work supports and enhances the development of suitable microgels for size-specific applications.
	The presented method efficiently explores new microgel synthesis recipes that facilitate tailor-made microgel production.
	
	\section*{Authors contributions}
	L.F.K.: conceptualization, methodology, Raman spectra evaluation, DLS measurement interpretation, designing experimental studies, TS-EMO and MAiNGO optimization configuration, graphic design, writing original draft;
	A.M.S.: TS-EMO optimization configuration, scientific support and discussion on interpretation of computational results, reviewing and editing the article;
	J.K.: assistance for experimental setup, synthesis conduction, Raman spectra acquisition, reviewing the article;
	J.I.: synthesis conduction, Raman spectra acquisition, conducting DLS measurements, reviewing the article;
	N.W.: conducting DLS measurements, reviewing and editing the article;
	A.M.: design of the project, scientific support and discussion on interpretation of computational results, and advice on the structure and presentation of this work, reviewing and editing the article.
	
	\section*{Acknowledgements}
	This work was performed as a part of project B4 of the CRC~985 ``Functional Microgels and Microgel Systems'' funded by Deutsche Forschungsgemeinschaft (DFG).
	The authors thank Jan Steinstraßen for support with conducting continuous microgel syntheses.
	The authors also thank Johannes M. M. Faust for fruitful discussions and Jannik Lüthje and Daniel Jungen for support with the software MAiNGO.
	
	\bibliographystyle{elsarticle-num}
	\bibliography{BIB4}

\begin{thebibliography}{10}
\expandafter\ifx\csname url\endcsname\relax
  \def\url#1{\texttt{#1}}\fi
\expandafter\ifx\csname urlprefix\endcsname\relax\def\urlprefix{URL }\fi
\expandafter\ifx\csname href\endcsname\relax
  \def\href#1#2{#2} \def\path#1{#1}\fi

\bibitem{Pich.2011}
A.~Pich, W.~Richtering, Chemical Design of Responsive Microgels, Vol. 234,
  {Springer Berlin Heidelberg}, Berlin, Heidelberg, 2011.
\newblock \href {https://doi.org/10.1007/978-3-642-16379-1}
  {\path{doi:10.1007/978-3-642-16379-1}}.

\bibitem{Keskin.2019}
D.~Keskin, O.~Mergel, H.~C. {van der Mei}, H.~J. Busscher, P.~{van Rijn},
  Inhibiting bacterial adhesion by mechanically modulated microgel coatings,
  Biomacromolecules 20~(1) (2019) 243--253.
\newblock \href {https://doi.org/10.1021/acs.biomac.8b01378}
  {\path{doi:10.1021/acs.biomac.8b01378}}.

\bibitem{Switacz.2020}
V.~K. Switacz, S.~K. Wypysek, R.~Degen, J.~J. Crassous, M.~Spehr,
  W.~Richtering, Influence of size and cross-linking density of microgels on
  cellular uptake and uptake kinetics, Biomacromolecules 21~(11) (2020)
  4532--4544.
\newblock \href {https://doi.org/10.1021/acs.biomac.0c00478}
  {\path{doi:10.1021/acs.biomac.0c00478}}.

\bibitem{Zhang.2019}
C.~Zhang, E.~Gau, W.~Sun, J.~Zhu, B.~M. Schmidt, A.~Pich, X.~Shi, Influence of
  size, crosslinking degree and surface structure of
  poly(n-vinylcaprolactam)-based microgels on their penetration into
  multicellular tumor spheroids, Biomaterials science 7~(11) (2019) 4738--4747.
\newblock \href {https://doi.org/10.1039/c9 bm01132c} {\path{doi:10.1039/c9
  bm01132c}}.

\bibitem{Faulde.2018}
M.~Faulde, E.~Siemes, D.~W{\"o}ll, A.~Jupke, Fluid dynamics of microgel-covered
  drops reveal impact on interfacial conditions, Polymers 10~(8) (2018).
\newblock \href {https://doi.org/10.3390/polym10080809}
  {\path{doi:10.3390/polym10080809}}.

\bibitem{Destribats.2014}
M.~Destribats, M.~Eyharts, V.~Lapeyre, E.~Sellier, I.~Varga, V.~Ravaine,
  V.~Schmitt, Impact of pnipam microgel size on its ability to stabilize
  pickering emulsions, Langmuir : the ACS journal of surfaces and colloids
  30~(7) (2014) 1768--1777.
\newblock \href {https://doi.org/10.1021/la4044396}
  {\path{doi:10.1021/la4044396}}.

\bibitem{Richtering.2012}
W.~Richtering, Responsive emulsions stabilized by stimuli-sensitive microgels:
  emulsions with special non-pickering properties, Langmuir : the ACS journal
  of surfaces and colloids 28~(50) (2012) 17218--17229.
\newblock \href {https://doi.org/10.1021/la302331s}
  {\path{doi:10.1021/la302331s}}.

\bibitem{Khan.2020}
S.~R. Khan, S.~Ali, B.~Ullah, S.~Jamil, T.~Zanib, Synthesis of iron
  nanoparticles in poly(n-isopropylacrylamide-acrylic acid) hybrid microgels
  for catalytic reduction of series of organic pollutants: a first approach,
  Journal of Nanoparticle Research 22~(7) (2020) 72.
\newblock \href {https://doi.org/10.1007/s11051-020-04924-5}
  {\path{doi:10.1007/s11051-020-04924-5}}.

\bibitem{Wolff.2018}
H.~J.~M. Wolff, M.~Kather, H.~Breisig, W.~Richtering, A.~Pich, M.~Wessling,
  From batch to continuous precipitation polymerization of thermoresponsive
  microgels, ACS applied materials {\&} interfaces 10~(29) (2018) 24799--24806.
\newblock \href {https://doi.org/10.1021/acsami.8b06920}
  {\path{doi:10.1021/acsami.8b06920}}.

\bibitem{Kather.2018}
M.~Kather, F.~Ritter, A.~Pich, Surfactant-free synthesis of extremely small
  stimuli-responsive colloidal gels using a confined impinging jet reactor,
  Chemical Engineering Journal 344 (2018) 375--379.
\newblock \href {https://doi.org/10.1016/j.cej.2018.03.082}
  {\path{doi:10.1016/j.cej.2018.03.082}}.

\bibitem{Fandrich.2020}
P.~Fandrich, L.~Wiehemeier, M.~Dirksen, O.~Wrede, T.~Kottke, T.~Hellweg,
  Acrylamide precipitation polymerization in a continuous flow reactor: an in
  situ ftir study reveals kinetics, Colloid and Polymer Science 299~(2) (2020)
  221--232.
\newblock \href {https://doi.org/10.1007/s00396-020-04762-w}
  {\path{doi:10.1007/s00396-020-04762-w}}.

\bibitem{Kaven.2021}
L.~Kaven, H.~Wolff, L.~Wille, M.~Wessling, A.~Mitsos, J.~Viell, In-line
  monitoring of microgel synthesis: Flow versus batch reactor, Organic process
  research {\&} development (Under Revision) (2021).

\bibitem{Fandrich.2023}
P.~Fandrich, J.~{Esteban V{\'a}zquez}, R.~Haverkamp, T.~Hellweg, Growth of
  smart microgels in a flow reactor scrutinized by in-line saxs, Langmuir : the
  ACS journal of surfaces and colloids (2023).
\newblock \href {https://doi.org/10.1021/acs.langmuir.2c02796}
  {\path{doi:10.1021/acs.langmuir.2c02796}}.

\bibitem{Janssen.2019}
F.~A.~L. Janssen, M.~Kather, A.~Ksiazkiewicz, A.~Pich, A.~Mitsos, Synthesis of
  poly(n-vinylcaprolactam)-based microgels by precipitation polymerization:
  Pseudo-bulk model for particle growth and size distribution, ACS omega 4~(9)
  (2019) 13795--13807.
\newblock \href {https://doi.org/10.1021/acsomega.9b01335}
  {\path{doi:10.1021/acsomega.9b01335}}.

\bibitem{Jung.2019}
F.~Jung, A.~Ksiazkiewicz, A.~Mhamdi, A.~Pich, A.~Mitsos, Model-based prediction
  of the hydrodynamic radius of collapsed microgels and experimental
  validation, Chemical Engineering Journal 378~(8) (2019) 121740.
\newblock \href {https://doi.org/10.1016/j.cej.2019.05.101}
  {\path{doi:10.1016/j.cej.2019.05.101}}.

\bibitem{Hoare.2006}
T.~Hoare, D.~McLean, Kinetic prediction of functional group distributions in
  thermosensitive microgels, The Journal of Physical Chemistry B 110~(41)
  (2006) 20327--20336.
\newblock \href {https://doi.org/10.1021/jp0643451}
  {\path{doi:10.1021/jp0643451}}.

\bibitem{Janssen.2017}
F.~A.~L. Janssen, M.~Kather, L.~C. Kr{\"o}ger, A.~Mhamdi, K.~Leonhard, A.~Pich,
  A.~Mitsos, Synthesis of poly( n -vinylcaprolactam)-based microgels by
  precipitation polymerization: Process modeling and experimental validation,
  Industrial {\&} Engineering Chemistry Research 56~(49) (2017) 14545--14556.
\newblock \href {https://doi.org/10.1021/acs.iecr.7b03263}
  {\path{doi:10.1021/acs.iecr.7b03263}}.

\bibitem{Janssen.2018}
F.~A. Janssen, A.~Ksiazkiewicz, M.~Kather, L.~C. Kr{\"o}ger, A.~Mhamdi,
  K.~Leonhard, A.~Pich, A.~Mitsos, Kinetic modeling of precipitation
  terpolymerization for functional microgels, in: 28th European Symposium on
  Computer Aided Process Engineering, Vol.~43 of Computer Aided Chemical
  Engineering, Elsevier, 2018, pp. 109--114.
\newblock \href {https://doi.org/10.1016/B978-0-444-64235-6.50021-8}
  {\path{doi:10.1016/B978-0-444-64235-6.50021-8}}.

\bibitem{Jung.2019b}
F.~Jung, A.~Ksiazkiewicz, A.~Mhamdi, A.~Pich, A.~Mitsos, Model-based prediction
  of the hydrodynamic radius of collapsed microgels and experimental
  validation, Chemical Engineering Journal 378 (2019) 121740.
\newblock \href {https://doi.org/10.1016/j.cej.2019.05.101}
  {\path{doi:10.1016/j.cej.2019.05.101}}.

\bibitem{Bradford.2018}
E.~Bradford, A.~M. Schweidtmann, A.~Lapkin, Efficient multiobjective
  optimization employing gaussian processes, spectral sampling and a genetic
  algorithm, Journal of Global Optimization 71~(2) (2018) 407--438.
\newblock \href {https://doi.org/10.1007/s10898-018-0609-2}
  {\path{doi:10.1007/s10898-018-0609-2}}.

\bibitem{Shields.2021}
B.~J. Shields, J.~Stevens, J.~Li, M.~Parasram, F.~Damani, J.~I.~M. Alvarado,
  J.~M. Janey, R.~P. Adams, A.~G. Doyle, Bayesian reaction optimization as a
  tool for chemical synthesis, Nature 590~(7844) (2021) 89--96.
\newblock \href {https://doi.org/10.1038/s41586-021-03213-y}
  {\path{doi:10.1038/s41586-021-03213-y}}.

\bibitem{Schweidtmann.2018}
A.~M. Schweidtmann, A.~D. Clayton, N.~Holmes, E.~Bradford, R.~A. Bourne, A.~A.
  Lapkin, Machine learning meets continuous flow chemistry: Automated
  optimization towards the pareto front of multiple objectives, Chemical
  Engineering Journal 352~(1--9) (2018) 277--282.
\newblock \href {https://doi.org/10.1016/j.cej.2018.07.031}
  {\path{doi:10.1016/j.cej.2018.07.031}}.

\bibitem{Sano.2020}
S.~Sano, T.~Kadowaki, K.~Tsuda, S.~Kimura, Application of bayesian optimization
  for pharmaceutical product development, Journal of Pharmaceutical Innovation
  15~(3) (2020) 333--343.
\newblock \href {https://doi.org/10.1007/s12247-019-09382-8}
  {\path{doi:10.1007/s12247-019-09382-8}}.

\bibitem{Naito.2022}
Y.~Naito, M.~Kondo, Y.~Nakamura, N.~Shida, K.~Ishikawa, T.~Washio, S.~Takizawa,
  M.~Atobe, Bayesian optimization with constraint on passed charge for
  multiparameter screening of electrochemical reductive carboxylation in a flow
  microreactor, Chemical communications (Cambridge, England) 58~(24) (2022)
  3893--3896.
\newblock \href {https://doi.org/10.1039/d2cc00124a}
  {\path{doi:10.1039/d2cc00124a}}.

\bibitem{Mogilicharla.2015}
A.~Mogilicharla, P.~Mittal, S.~Majumdar, K.~Mitra, Kriging surrogate based
  multi-objective optimization of bulk vinyl acetate polymerization with
  branching, Materials and Manufacturing Processes 30~(4) (2015) 394--402.
\newblock \href {https://doi.org/10.1080/10426914.2014.921709}
  {\path{doi:10.1080/10426914.2014.921709}}.

\bibitem{McPhee.1993}
W.~McPhee, K.~C. Tam, R.~Pelton, Poly(n-isopropylacrylamide) latices prepared
  with sodium dodecyl sulfate, Journal of Colloid and Interface Science 156~(1)
  (1993) 24--30.
\newblock \href {https://doi.org/10.1006/jcis.1993.1075}
  {\path{doi:10.1006/jcis.1993.1075}}.

\bibitem{Wu.1994}
X.~Wu, R.~H. Pelton, A.~E. Hamielec, D.~R. Woods, W.~McPhee, The kinetics of
  poly(n-isopropylacrylamide) microgel latex formation, Colloid and Polymer
  Science 272~(4) (1994) 467--477.
\newblock \href {https://doi.org/10.1007/BF00659460}
  {\path{doi:10.1007/BF00659460}}.

\bibitem{Andersson.2006}
M.~Andersson, S.~L. Maunu, Structural studies of poly(n-isopropylacrylamide)
  microgels: Effect of sds surfactant concentration in the microgel synthesis,
  Journal of Polymer Science Part B: Polymer Physics 44~(23) (2006) 3305--3314.
\newblock \href {https://doi.org/10.1002/polb.20971}
  {\path{doi:10.1002/polb.20971}}.

\bibitem{Wedel.2017}
B.~Wedel, T.~Br{\"a}ndel, J.~Bookhold, T.~Hellweg, Role of anionic surfactants
  in the synthesis of smart microgels based on different acrylamides, ACS omega
  2~(1) (2017) 84--90.
\newblock \href {https://doi.org/10.1021/acsomega.6b00424}
  {\path{doi:10.1021/acsomega.6b00424}}.

\bibitem{Nessen.2013}
K.~von Nessen, M.~Karg, T.~Hellweg, Thermoresponsive
  poly-(n-isopropylmethacrylamide) microgels: Tailoring particle size by
  interfacial tension control, Polymer 54~(21) (2013) 5499--5510.
\newblock \href {https://doi.org/10.1016/j.polymer.2013.08.027}
  {\path{doi:10.1016/j.polymer.2013.08.027}}.

\bibitem{Virtanen.2014}
O.~L.~J. Virtanen, W.~Richtering, Kinetics and particle size control in
  non-stirred precipitation polymerization of n-isopropylacrylamide, Colloid
  and Polymer Science 292~(8) (2014) 1743--1756.
\newblock \href {https://doi.org/10.1007/s00396-014-3208-x}
  {\path{doi:10.1007/s00396-014-3208-x}}.

\bibitem{Balaceanu.2011}
A.~Balaceanu, D.~E. Demco, M.~M{\"o}ller, A.~Pich, Microgel heterogeneous
  morphology reflected in temperature-induced volume transition and 1 h
  high-resolution transverse relaxation nmr. the case of poly( n
  -vinylcaprolactam) microgel, Macromolecules 44~(7) (2011) 2161--2169.
\newblock \href {https://doi.org/10.1021/ma200103y}
  {\path{doi:10.1021/ma200103y}}.

\bibitem{Schneider.2014}
F.~Schneider, A.~Balaceanu, A.~Feoktystov, V.~Pipich, Y.~Wu, J.~Allgaier,
  W.~Pyckhout-Hintzen, A.~Pich, G.~J. Schneider, Monitoring the internal
  structure of poly(n-vinylcaprolactam) microgels with variable cross-link
  concentration, Langmuir : the ACS journal of surfaces and colloids 30~(50)
  (2014) 15317--15326.
\newblock \href {https://doi.org/10.1021/la503830w}
  {\path{doi:10.1021/la503830w}}.

\bibitem{Virtanen.2019}
O.~L.~J. Virtanen, M.~Kather, J.~Meyer-Kirschner, A.~Melle, A.~Radulescu,
  J.~Viell, A.~Mitsos, A.~Pich, W.~Richtering, Direct monitoring of microgel
  formation during precipitation polymerization of n-isopropylacrylamide using
  in situ sans, ACS omega 4~(2) (2019) 3690--3699.
\newblock \href {https://doi.org/10.1021/acsomega.8b03461}
  {\path{doi:10.1021/acsomega.8b03461}}.

\bibitem{Imaz.2008}
A.~Imaz, J.~Forcada, N -vinylcaprolactam-based microgels: Synthesis and
  characterization, Journal of Polymer Science Part A: Polymer Chemistry 46~(7)
  (2008) 2510--2524.
\newblock \href {https://doi.org/10.1002/pola.22583}
  {\path{doi:10.1002/pola.22583}}.

\bibitem{Chiu.1995}
Y.~Y. Chiu, L.~J. Lee, Microgel formation in the free radical crosslinking
  polymerization of ethylene glycol dimethacrylate (egdma). i. experimental,
  Journal of Polymer Science Part A: Polymer Chemistry 33~(2) (1995) 257--267.
\newblock \href {https://doi.org/10.1002/pola.1995.080330208}
  {\path{doi:10.1002/pola.1995.080330208}}.

\bibitem{Bongartz.2018}
D.~Bongartz, J.~Najman, S.~Sass, A.~Mitsos,
  \href{https://www.avt.rwth-aachen.de/global/show_document.asp?id=aaaaaaaaabclahw}{Maingo
  - mccormick-based algorithm for mixed-integer nonlinear global optimization.
  technical report}.
\newline\urlprefix\url{https://www.avt.rwth-aachen.de/global/show_document.asp?id=aaaaaaaaabclahw}

\bibitem{Schweidtmann.2021}
A.~M. Schweidtmann, D.~Bongartz, D.~Grothe, T.~Kerkenhoff, X.~Lin, J.~Najman,
  A.~Mitsos, Deterministic global optimization with gaussian processes
  embedded, Mathematical Programming Computation 13~(3) (2021) 553--581.
\newblock \href {https://doi.org/10.1007/s12532-021-00204-y}
  {\path{doi:10.1007/s12532-021-00204-y}}.

\bibitem{Kriesten.2008}
E.~Kriesten, F.~Alsmeyer, A.~Bardow, W.~Marquardt, Fully automated indirect
  hard modeling of mixture spectra, Chemometrics and Intelligent Laboratory
  Systems 91~(2) (2008) 181--193.
\newblock \href {https://doi.org/10.1016/j.chemolab.2007.11.004}
  {\path{doi:10.1016/j.chemolab.2007.11.004}}.

\bibitem{data.Kaven.2021}
L.~Kaven, H.~Wolff, L.~Wille, M.~Wessling, A.~Mitsos, J.~Viell, Dataset to:
  In-line monitoring of microgel synthesis: Flow versus batch reactor,
  doi:\url{10.18154/RWTH-2021-09666 } (2021).

\bibitem{SigmaAldrich.2023}
{Sigma-Aldrich Chemie GmbH},
  \href{https://www.sigmaaldrich.com/DE/en/sds/aldrich/440914}{Safety data
  sheet 440914} version 7.2 (2023).
\newline\urlprefix\url{https://www.sigmaaldrich.com/DE/en/sds/aldrich/440914}

\bibitem{Ehrgott.2009}
M.~Ehrgott, Multiobjective optimization, AI Magazine 29~(4) (2009) 47.
\newblock \href {https://doi.org/10.1609/aimag.v29i4.2198}
  {\path{doi:10.1609/aimag.v29i4.2198}}.

\bibitem{data.Kaven.2023}
L.~Kaven, A.~M. Schweidtmann, J.~Keil, J.~Israel, N.~Wolter, A.~Mitsos, Dataset
  to: Data-driven product-process optimization of n-isopropylacrylamide
  microgel flow-synthesis, doi:\url{10.18154/RWTH-2023-05551} (2023).

\bibitem{tsemo}
E.~Bradford, {TS-EMO algorithm}, \url{https://github.com/Eric-Bradford/TS-EMO},
  last updated 2020-06.

\bibitem{maingo}
D.~Bongartz, {MAiNGO - McCormick-based Algorithm for mixed-integer Nonlinear
  Global Optimization}, \url{https://git.rwth-aachen.de/avt-svt/public/maingo},
  last updated 2021-06.

\bibitem{melon}
A.~Schweidtmann, {MeLOn - Machine Learning Models for Optimization},
  \url{https://git.rwth-aachen.de/avt-svt/public/MeLOn}, last updated 2021-06.

\end{thebibliography}

\end{document}


\title{Supporting Information: Data-driven Product-Process Optimization of N-isopropylacrylamide Microgel Synthesis in Flow}
	\author[1]{Luise F. Kaven}
	\author[2]{Artur M. Schweidtmann}
	\author[1]{Jan Keil}
	\author[1]{Jana Israel}
	\author[3,4]{Nadja Wolter}
	\author[1,5,*]{Alexander Mitsos}
	
	\affil[1]{Process Systems Engineering (AVT.SVT), RWTH Aachen University, Forckenbeckstr. 51, 52074 Aachen, Germany}
	\affil[2]{Department of Chemical Engineering, Delft University of Technology, Van der Maasweg 9, 2629 HZ Delft, The Netherlands}
	\affil[3]{DWI - Leibniz Institute for Interactive Materials e.V., Forckenbeckstr. 50, 52074 Aachen, Germany}
	\affil[4]{Functional and Interactive Polymers, Institute of Technical and Macromolecular Chemistry, RWTH Aachen University, Forckenbeckstr. 50, 52074 Aachen, Germany}
	\affil[5]{JARA-soft, RWTH Aachen University, Templergraben 55, 52056 Aachen, Germany}
	\affil[*]{Corresponding author, E-mail: amitsos@alum.mit.edu, Forckenbeckstr. 51, 52074 Aachen, Germany}
	
	\date{June 2023}
	
	\maketitle

	\section{Gaussian processes}\label{sec:GPs}
	
	\begin{figure*}[h]
		\centering
		\subfigure[]
		{
			\label{subfig:GP_product_4inputs3obj_Temp}
			\includegraphics[width=0.45\textwidth]{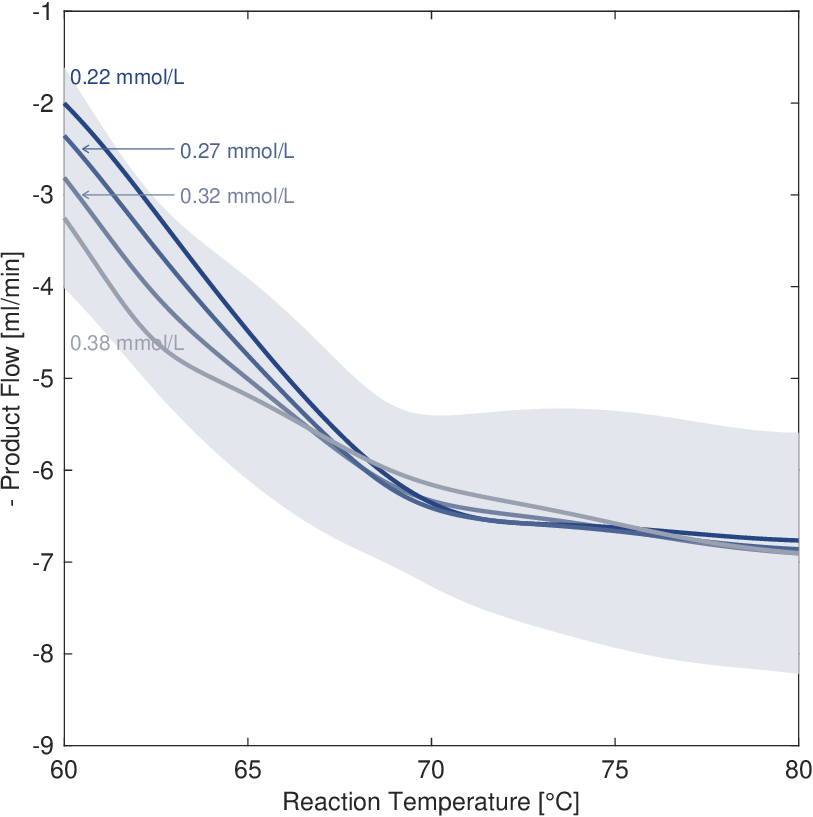}
		}
		\subfigure[]
		{
			\label{subfig:GP_product_4inputs3obj_Initiator}
			\includegraphics[width=0.45\textwidth]{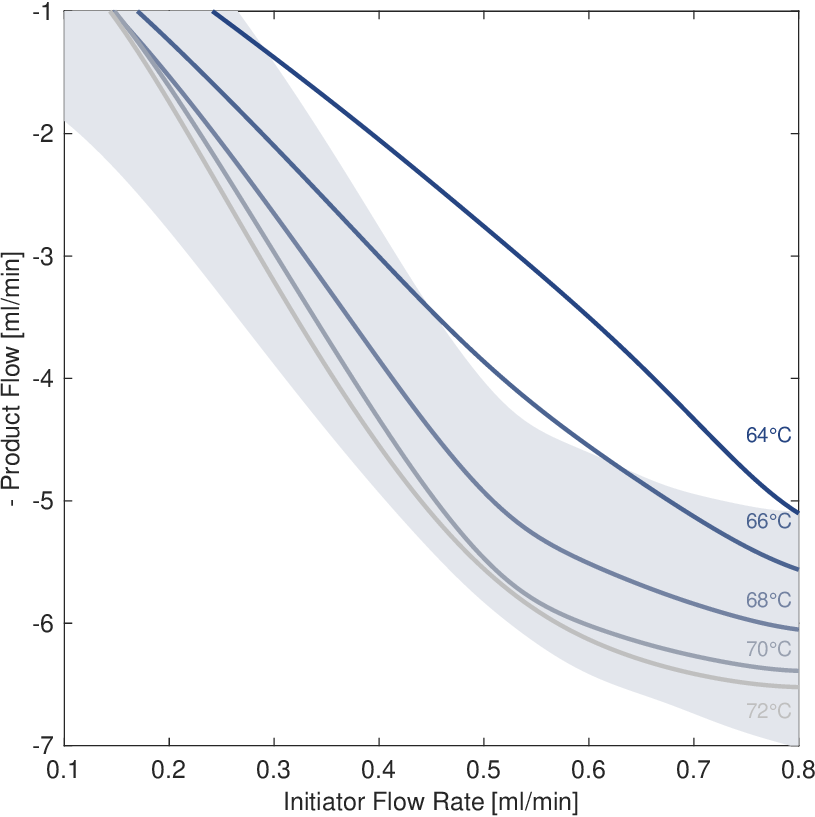}
		}\\
		\subfigure[]
		{
			\includegraphics[width=0.45\textwidth]{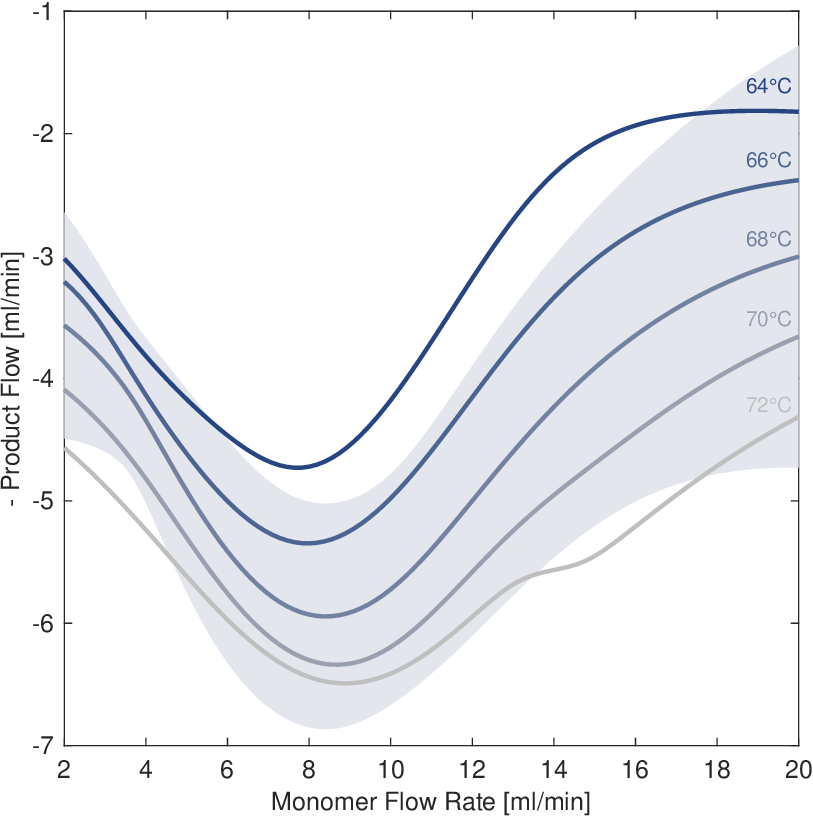}
			\label{subfig:GP_product_4inputs3obj_Monomer}
		}
		\subfigure[]
		{
			\includegraphics[width=0.45\textwidth]{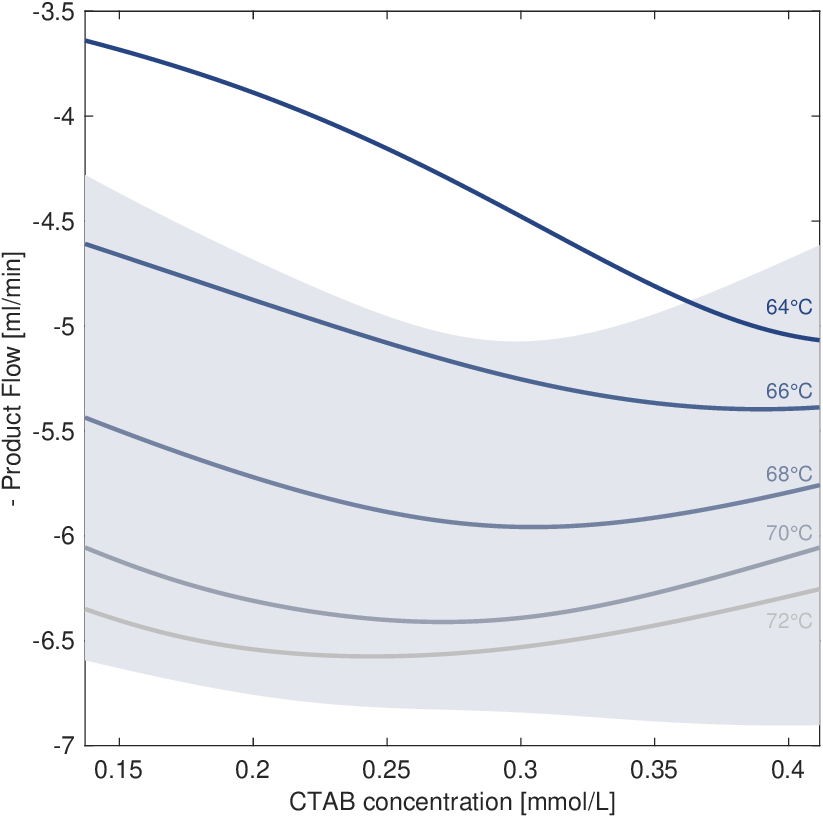}
			\label{subfig:GP_product_4inputs3obj_CTAB}
		}
		\caption{GP prediction of product flow as a function of: (a) reaction temperature, (b) initiator flow rate, (c) monomer flow rate, and (d) surfactant concentration.
			The GP is trained on data points generated in the hardware-in-the-loop study involving TS-EMO.
			The shaded are represents the variance of the prediction at \SI{0.32}{\milli\mole\per\liter} for (a).
			For (b), (c), and (d), the shaded area denotes the variance of the prediction at \SI{68}{\degreeCelsius}.}
		\label{fig:GP_4inputs3obj_Flow}%
	\end{figure*}
	
	\FloatBarrier
	
	\begin{figure*}[h]
		\centering
		\subfigure[]
		{
			\label{subfig:GP_size_4inputs3obj_Temp}
			\includegraphics[width=0.45\textwidth]{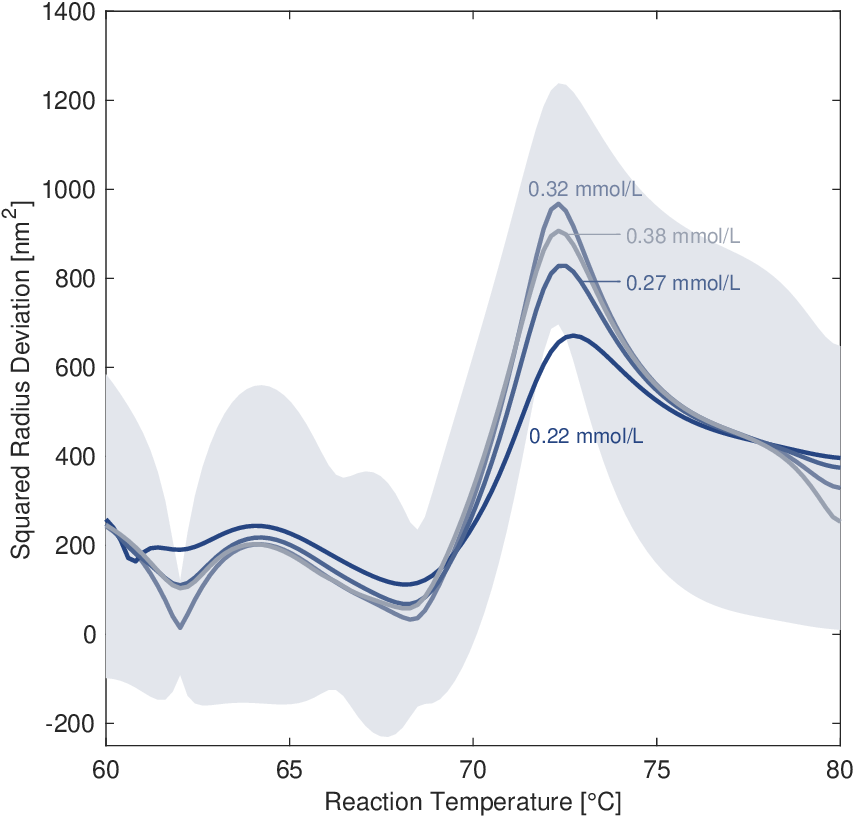}
		}
		\subfigure[]
		{
			\label{subfig:GP_size_4inputs3obj_Initiator}
			\includegraphics[width=0.45\textwidth]{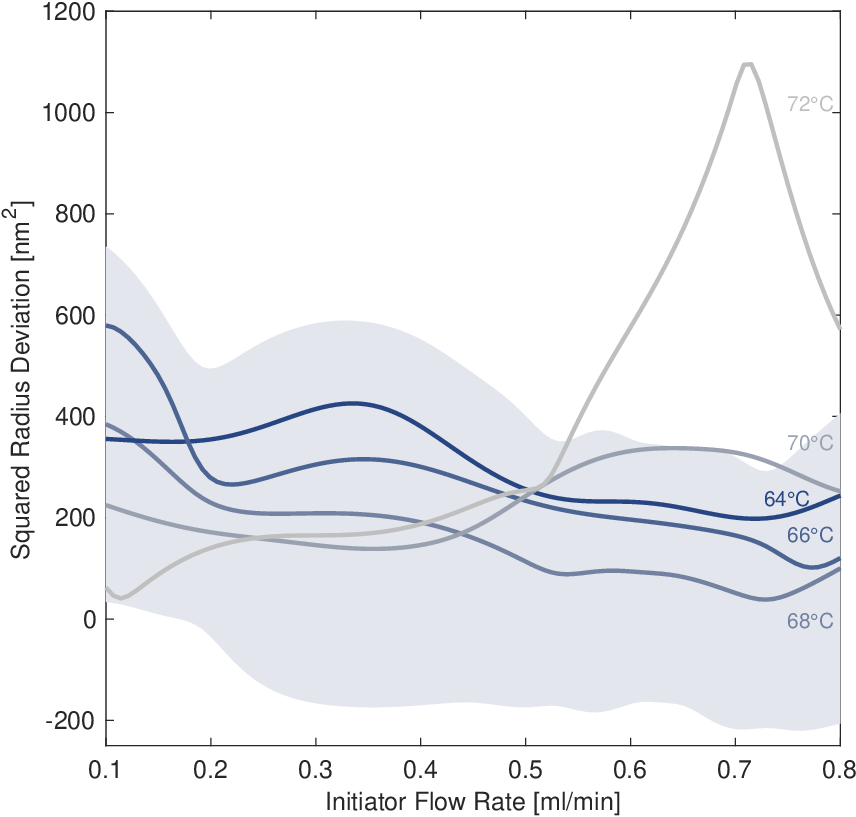}
		}\\
		\subfigure[]
		{
			\includegraphics[width=0.45\textwidth]{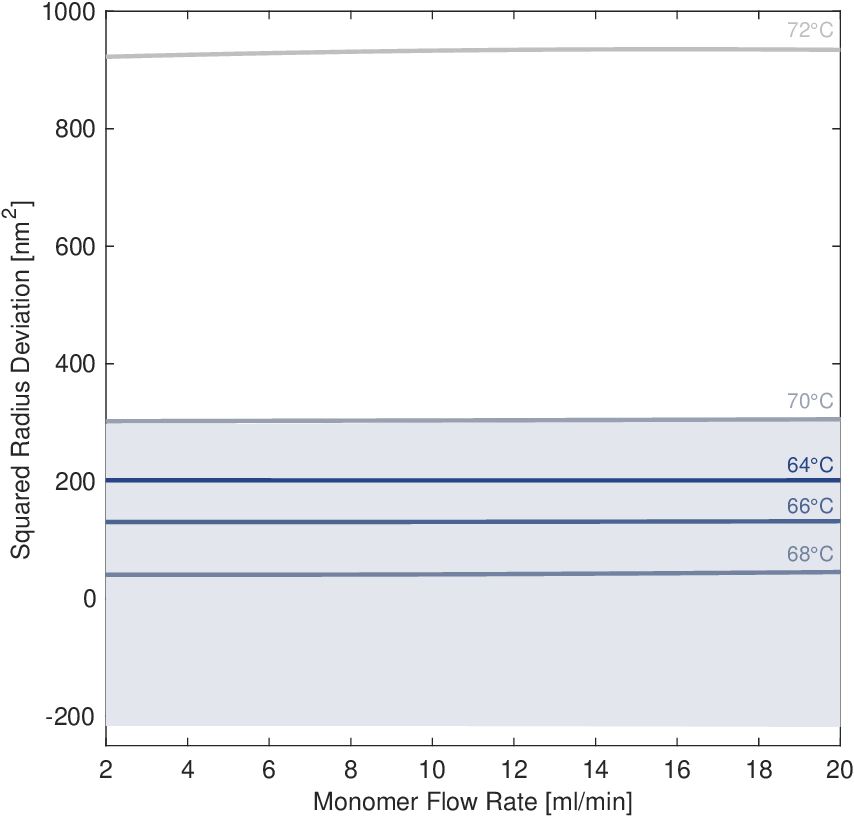}
			\label{subfig:GP_size_4inputs3obj_Monomer}
		}
		\subfigure[]
		{
			\includegraphics[width=0.45\textwidth]{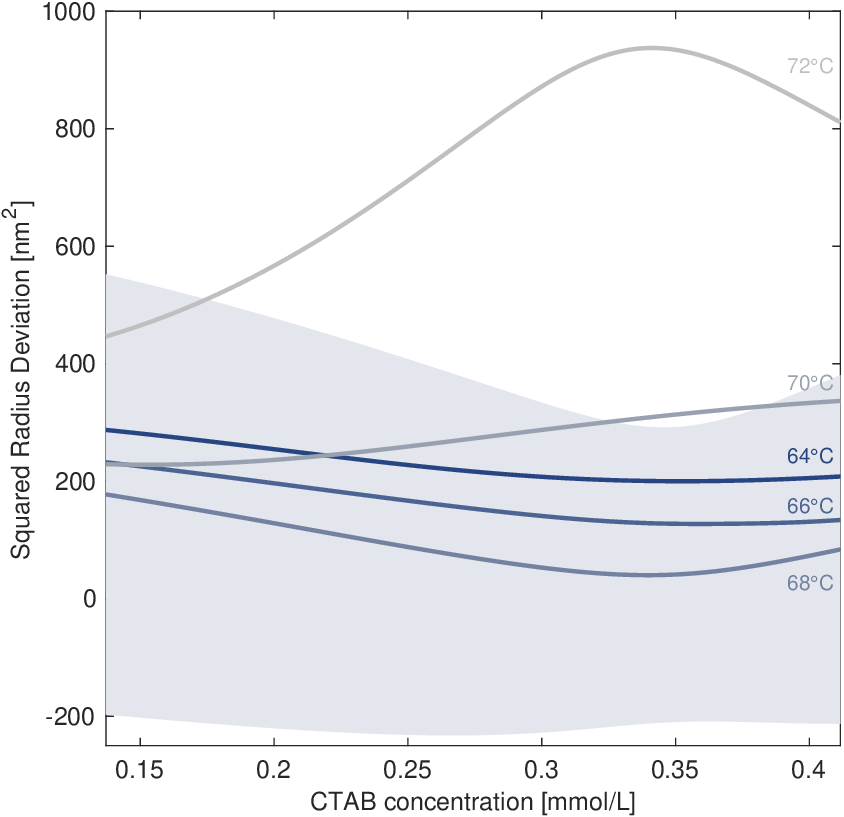}
			\label{subfig:GP_size_4inputs3obj_CTAB}
		}
		\caption{GP prediction of squared radius deviation as a function of: (a) reaction temperature, (b) initiator flow rate, (c) monomer flow rate, and (d) surfactant concentration.
			The GP is trained on data points generated in the hardware-in-the-loop study involving TS-EMO.
			The shaded are represents the variance of the prediction at \SI{0.32}{\milli\mole\per\liter} for (a).
			For (b), (c), and (d), the shaded area denotes the variance of the prediction at \SI{68}{\degreeCelsius}.
		}
		\label{fig:GP_4inputs3obj_Size}%
	\end{figure*}
	\FloatBarrier
	
	\begin{figure*}[h]
		\centering
		\subfigure[]
		{
			\label{subfig:GP_temp_4inputs3obj_Temp}
			\includegraphics[width=0.45\textwidth]{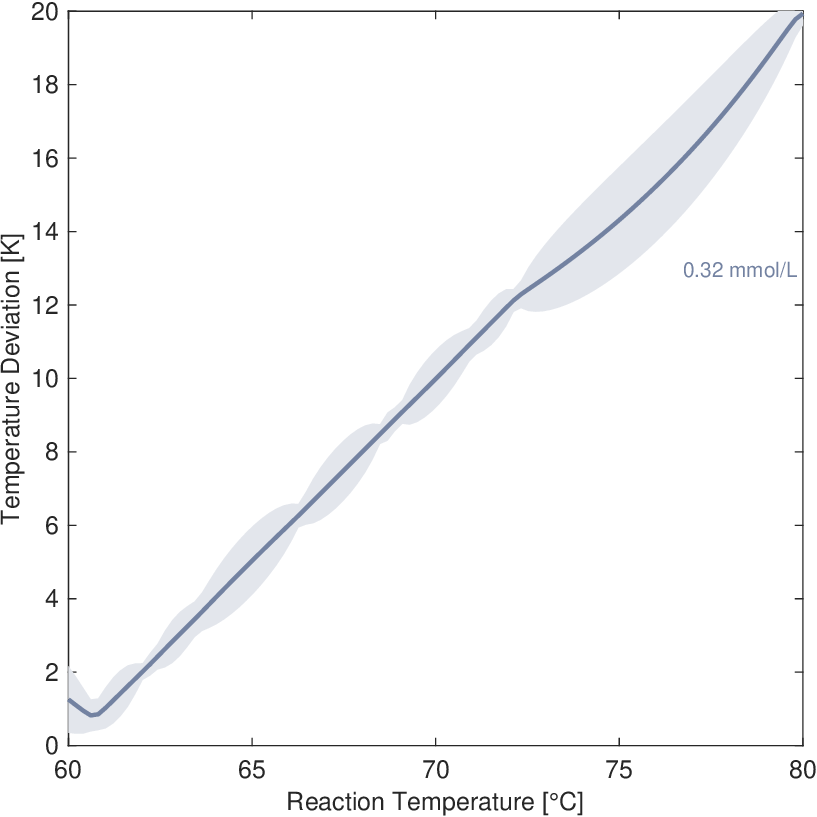}
		}
		\subfigure[]
		{
			\label{subfig:GP_temp_4inputs3obj_Initiator}
			\includegraphics[width=0.45\textwidth]{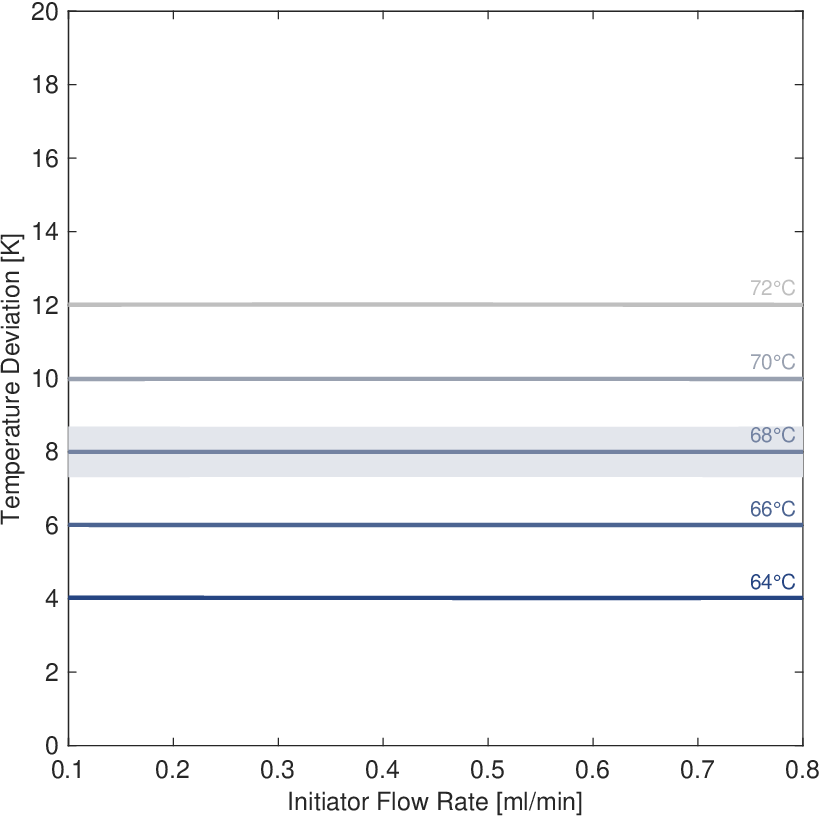}
		}\\
		\subfigure[]
		{
			\includegraphics[width=0.45\textwidth]{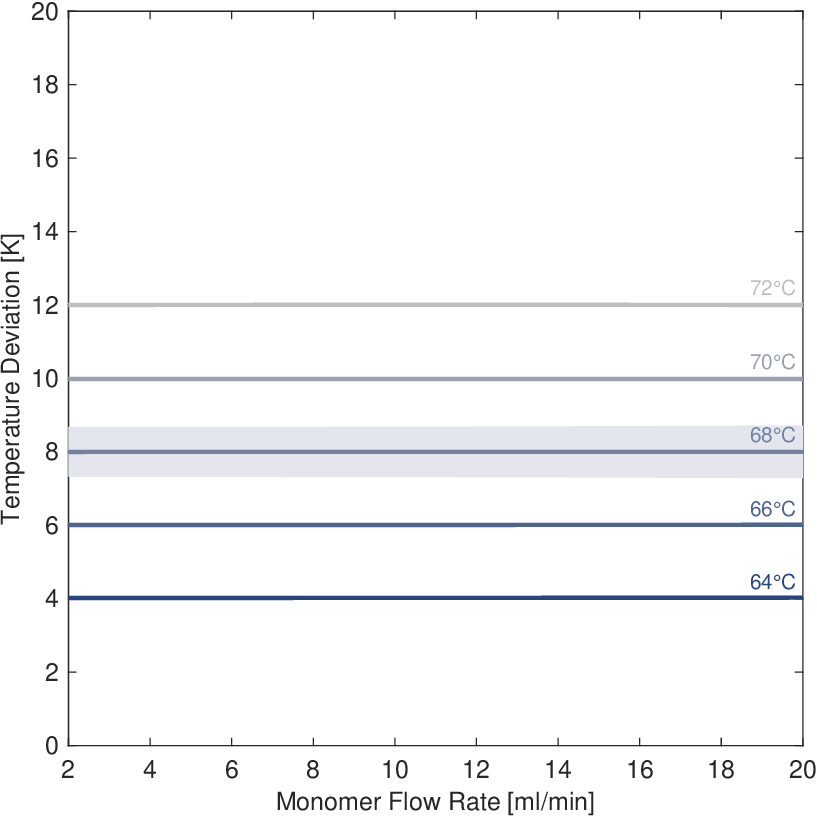}
			\label{subfig:GP_temp_4inputs3obj_Monomer}
		}
		\subfigure[]
		{
			\includegraphics[width=0.45\textwidth]{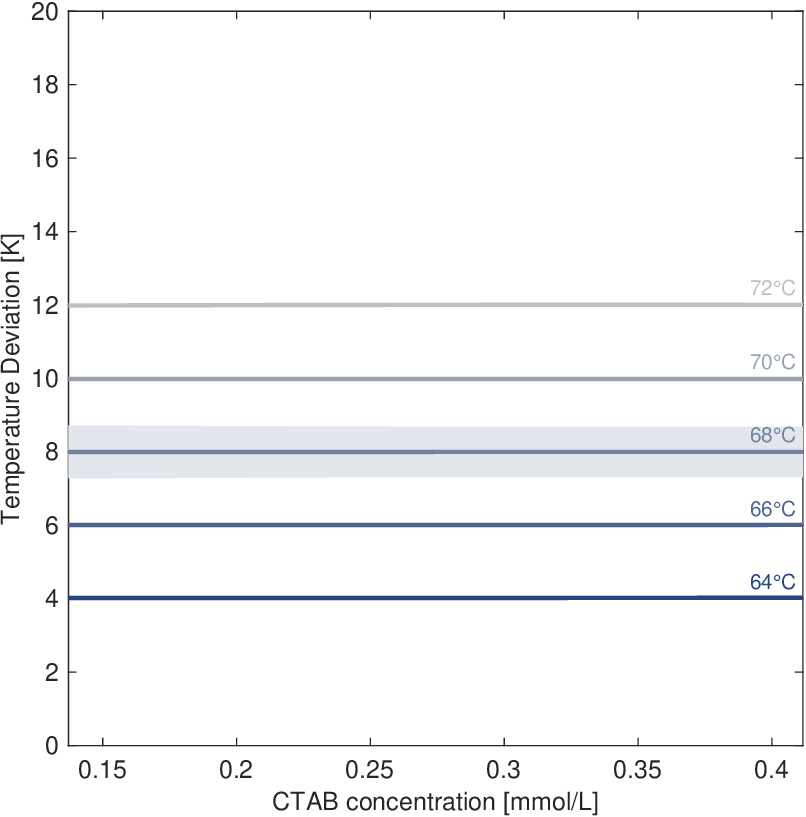}
			\label{subfig:GP_temp_4inputs3obj_CTAB}
		}
		\caption{GP prediction of temperature deviation as a function of: (a) reaction temperature, (b) initiator flow rate, (c) monomer flow rate, and (d) surfactant concentration.
			The GP is trained on data points generated in the hardware-in-the-loop study involving TS-EMO.
			The shaded are represents the variance of the prediction at \SI{0.32}{\milli\mole\per\liter} for (a).
			For (b), (c), and (d), the shaded area denotes the variance of the prediction at \SI{68}{\degreeCelsius}.}
		\label{fig:GP_4inputs3obj_Temp}%
	\end{figure*}
	\FloatBarrier

	\section{Data tables}\label{sec:data_tables}
	
	\subsection{Data regarding hardware-in-the-loop study involving TS-EMO} \label{sec:si_datatsemo}
	
	The data presented in Tab.~\ref{tbl:data_TSEMO} enables the reconstruction of Figs~3, 4, and 6.
	
	\begin{table}[ht]%
		\caption[Table]{Hardware-in-the-loop input and output data.}
		\label{tbl:data_TSEMO}
		\centering
		\begin{tabular}{|c|cccc|cc|} 
			\hline
			\# & v\textsubscript{I} & v\textsubscript{M} & T & c\textsubscript{CTAB} & $\Delta$r$^2$\textsubscript{H} & F\textsubscript{product} \\ 
			
			iteration & [\SI{}{\milli\liter\per\minute}] & [\SI{}{\milli\liter\per\minute}] & [\SI{}{\degreeCelsius}] & [\SI{}{\milli\mole\per\liter}] & [\SI{}{\nano\meter\squared}] & [\SI{}{\milli\liter\per\minute}]\\
			
			\hline
			0  & 0.34  & 4.86 & 71  & 0.16  & 9.00  & -4.29\\
			0  & 0.46  & 10.46  & 71  & 0.16  & 169.00  & -4.53\\
			0  & 0.8  & 7.80 & 71  & 0.16  & 42.25  & -6.54\\
			0  & 0.52  & 16.42  & 71  & 0.16  & 1,056.25  & -2.49\\
			0  & 0.44  & 9.45 & 80  & 0.41  & 1,225.00  & -7.41\\
			0  & 0.59  & 13.58  & 80  & 0.41  & 1,332.25  & -8.75\\
			0  & 0.21  & 4.04 & 80  & 0.41  & 729.00  & -3.85\\
			0  & 0.73  & 5.23  & 80  & 0.41  & 210.25  & -5.50 \\
			0  & 0.74  & 3.67  & 62  & 0.33  & 1.00  & -3.43  \\
			0  & 0.65  & 13.76  & 62  & 0.33  & 90.25  & -0.20  \\
			0  & 0.11  & 5.51  & 62  & 0.33  & 110.25  & 0.00  \\
			1  & 0.70  & 13.12  & 72.2  & 0.34  & 1089.00  & -5.42  \\
			1  & 0.11  & 6.51  & 72.2  & 0.34  & 6.25  & -0.40  \\
			1  & 0.71  & 14.77  & 72.2  & 0.34  & 1260.25  & -5.52  \\
			1  & 0.52  & 5.14  & 72.2  & 0.34  & 225.00  & -5.06  \\
			1  & 0.10  & 14.03  & 72.2  & 0.34  & 64.00  & 0.00  \\
			2  & 0.77  & 2.60  & 66.3  & 0.37  & 30.25  & -3.26  \\
			2  & 0.18  & 3.90  & 66.3  & 0.37  & 182.25  & -2.83  \\
			2  & 0.73  & 3.80  & 66.3  & 0.37  & 144.00  & -4.13  \\
			2  & 0.10  & 4.00  & 66.3  & 0.37  & 676.00  & -1.56  \\
			2  & 0.17  & 3.30  & 66.3  & 0.37  & 484.00  & -2.84  \\
			3  & 0.36  & 2.02  & 63.5  & 0.14  & 900.00  & -2.15  \\
			3  & 0.58  & 5.96  & 63.5  & 0.14  & 289.00  & -3.64  \\
			3  & 0.59  & 3.40  & 63.5  & 0.14  & 600.25  & -3.39  \\
			3  & 0.56  & 10.64  & 63.5  & 0.14  & 132.25  & -1.40  \\
			3  & 0.56  & 6.24  & 63.5  & 0.14  & 210.25  & -4.25  \\
			4  & 0.80  & 9.91  & 69  & 0.29  & 121.00  & -6.30  \\
			4  & 0.67  & 7.52  & 69  & 0.29  & 90.25  & -6.13  \\
			4  & 0.46  & 6.97  & 69  & 0.29  & 144.00  & -5.25  \\
			4  & 0.53  & 7.98  & 69  & 0.29  & 144.00  & -5.88  \\
			5  & 0.44  & 3.58  & 62.4  & 0.21  & 196.00  & -0.71  \\
			5  & 0.44  & 2.94  & 62.4  & 0.21  & 240.25  & -0.49  \\
			6  & 0.77  & 6.70  & 62.3  & 0.41  & 210.25  & -4.54  \\
			6  & 0.80  & 7.52  & 62.3  & 0.41  & 462.25  & -6.14  \\
			7  & 0.61  & 5.53  & 68.5  & 0.35  & 81.00  & -5.16  \\
			7  & 0.65  & 4.22  & 68.5  & 0.35  & 90.25  & -4.46  \\
			7  & 0.19  & 2.68  & 68.5  & 0.35  & 196.00  & -2.68  \\
			7  & 0.73  & 3.63  & 68.5  & 0.35  & 12.25  & -4.08  \\
			7  & 0.53  & 3.06  & 68.5  & 0.35  & 42.25  & -3.43  \\
			8  & 0.70  & 4.22  & 60.7  & 0.23  & 484.00  & -0.54  \\
			8  & 0.60  & 2.66  & 60.7  & 0.23  & 16.00  & -1.26  \\
			8  & 0.74  & 3.67  & 60.7  & 0.23  & 132.25  & -0.90  \\
			8  & 0.70  & 3.03  & 60.7  & 0.23  & 529.00  & -0.20  \\
			\hline
		\end{tabular}
	\end{table}
	\FloatBarrier
	
	\subsection{Data regarding global deterministic optimization including MAiNGO} \label{sec:si_datamaingo}
	
	The data presented in Tabs.~\ref{tbl:data_MAINGO_80} to \ref{tbl:data_MAINGO_61} enables the reconstruction of Figs~7 and 8.
	
	\begin{table}[ht]%
		\caption[Table]{Pareto optimal solutions calculated via MAiNGO for a upper bound on the input temperature of \SI{80}{\degreeCelsius}.}
		\label{tbl:data_MAINGO_80}
		\centering
		\begin{tabular}{|cccc|cc|} 
			\hline
			v\textsubscript{I} & v\textsubscript{M} & T & c\textsubscript{CTAB} & $\Delta$r$^2$\textsubscript{H} & F\textsubscript{product} \\ 
			
			[\SI{}{\milli\liter\per\minute}] & [\SI{}{\milli\liter\per\minute}] & [\SI{}{\degreeCelsius}] & [\SI{}{\milli\mole\per\liter}] & [\SI{}{\nano\meter\squared}] & [\SI{}{\milli\liter\per\minute}]\\
			
			\hline
			0.73  & 8.44  & 68.52  & 0.34  & 25  & -6.04  \\
			0.73  & 8.44  & 68.51  & 0.34  & 24  & -6.04  \\
			0.73  & 8.44  & 68.50  & 0.34  & 23  & -6.04  \\
			0.73  & 8.44  & 68.50  & 0.34  & 22  & -6.03  \\
			0.73  & 8.43  & 68.49  & 0.34  & 21  & -6.03  \\
			0.73  & 8.43  & 68.49  & 0.34  & 20  & -6.02  \\
			0.73  & 8.42  & 68.48  & 0.34  & 19  & -6.02  \\
			0.73  & 8.38  & 68.46  & 0.35  & 18  & -6.01  \\
			0.73  & 7.69  & 68.46  & 0.35  & 17  & -5.95  \\
			0.73  & 6.84  & 68.47  & 0.35  & 16  & -5.76  \\
			0.73  & 5.98  & 68.48  & 0.35  & 15  & -5.44  \\
			0.73  & 5.13  & 68.49  & 0.35  & 14  & -5.01  \\
			0.73  & 4.27  & 68.49  & 0.35  & 13  & -4.50  \\
			0.34  & 4.88  & 71.00  & 0.16  & 12  & -4.30  \\
			0.34  & 4.87  & 71.00  & 0.16  & 11  & -4.30  \\
			0.34  & 4.87  & 71.00  & 0.16  & 10  & -4.29  \\
			0.74  & 6.63  & 62.00  & 0.33  & 9  & -3.78  \\
			0.74  & 6.26  & 62.00  & 0.33  & 8  & -3.70  \\
			0.74  & 5.88  & 62.00  & 0.33  & 7  & -3.62  \\
			0.74  & 5.51  & 62.00  & 0.33  & 6  & -3.54  \\
			0.74  & 5.14  & 62.00  & 0.33  & 5  & -3.48  \\
			0.74  & 3.70  & 62.01  & 0.33  & 4  & -3.44  \\
			0.74  & 3.69  & 62.01  & 0.33  & 3  & -3.43  \\
			0.74  & 3.68  & 62.00  & 0.33  & 2  & -3.43  \\
			\hline
		\end{tabular}
	\end{table}
	\FloatBarrier
	
	\begin{table}[ht]%
		\caption[Table]{Pareto optimal solutions calculated via MAiNGO for a upper bound on the input temperature of \SI{70}{\degreeCelsius}.}
		\label{tbl:data_MAINGO_70}
		\centering
		\begin{tabular}{|cccc|cc|} 
			\hline
			v\textsubscript{I} & v\textsubscript{M} & T & c\textsubscript{CTAB} & $\Delta$r$^2$\textsubscript{H} & F\textsubscript{product} \\ 
			
			[\SI{}{\milli\liter\per\minute}] & [\SI{}{\milli\liter\per\minute}] & [\SI{}{\degreeCelsius}] & [\SI{}{\milli\mole\per\liter}] & [\SI{}{\nano\meter\squared}] & [\SI{}{\milli\liter\per\minute}]\\
			
			\hline
			0.73  & 8.45  & 68.51  & 0.34  & 25  & -6.04  \\
			0.73  & 8.44  & 68.51  & 0.34  & 24  & -6.04  \\
			0.73  & 8.44  & 68.51  & 0.34  & 23  & -6.04  \\
			0.73  & 8.44  & 68.50  & 0.34  & 22  & -6.03  \\
			0.73  & 8.44  & 68.50  & 0.34  & 21  & -6.03  \\
			0.73  & 8.43  & 68.49  & 0.34  & 20  & -6.02  \\
			0.73  & 8.42  & 68.48  & 0.34  & 19  & -6.02  \\
			0.73  & 8.36  & 68.47  & 0.35  & 18  & -6.01  \\
			0.73  & 7.69  & 68.46  & 0.35  & 17  & -5.95  \\
			0.73  & 6.84  & 68.47  & 0.35  & 16  & -5.76  \\
			0.73  & 5.98  & 68.48  & 0.35  & 15  & -5.44  \\
			0.73  & 5.13  & 68.49  & 0.35  & 14  & -5.01  \\
			0.73  & 4.27  & 68.49  & 0.35  & 13  & -4.50  \\
			0.74  & 7.57  & 62.01  & 0.33  & 12  & -3.92  \\
			0.74  & 7.35  & 62.00  & 0.33  & 11  & -3.90  \\
			0.74  & 7.00  & 62.00  & 0.33  & 10  & -3.85  \\
			0.74  & 6.63  & 62.00  & 0.33  & 9  & -3.78  \\
			0.74  & 6.26  & 62.00  & 0.33  & 8  & -3.70  \\
			0.74  & 5.88  & 62.00  & 0.33  & 7  & -3.62  \\
			0.74  & 5.51  & 62.00  & 0.33  & 6  & -3.54  \\
			0.74  & 5.14  & 62.00  & 0.33  & 5  & -3.48  \\
			0.74  & 3.70  & 62.01  & 0.33 & 4  & -3.44  \\ 
			0.74  & 3.69  & 62.01  & 0.33  & 3  & -3.43  \\
			0.74  & 3.68  & 62.00  & 0.33  & 2  & -3.43  \\
			\hline
		\end{tabular}
	\end{table}
	\FloatBarrier

	\begin{table}[ht]%
		\caption[Table]{Pareto optimal solutions calculated via MAiNGO for a upper bound on the input temperature of \SI{62}{\degreeCelsius}.}
		\label{tbl:data_MAINGO_62}
		\centering
		\begin{tabular}{|cccc|cc|} 
			\hline
			v\textsubscript{I} & v\textsubscript{M} & T & c\textsubscript{CTAB} & $\Delta$r$^2$\textsubscript{H} & F\textsubscript{product} \\ 
			
			[\SI{}{\milli\liter\per\minute}] & [\SI{}{\milli\liter\per\minute}] & [\SI{}{\degreeCelsius}] & [\SI{}{\milli\mole\per\liter}] & [\SI{}{\nano\meter\squared}] & [\SI{}{\milli\liter\per\minute}]\\
			
			\hline
			0.74  & 7.78  & 62.00  & 0.34  & 25  & -4.02  \\
			0.74  & 7.77  & 62.00  & 0.34  & 24  & -4.02  \\
			0.74  & 7.77  & 62.00  & 0.34  & 23  & -4.01  \\
			0.74  & 7.77  & 62.00  & 0.34  & 22  & -4.00  \\
			0.74  & 7.76  & 62.00  & 0.34  & 21  & -4.00  \\
			0.74  & 7.76  & 62.00  & 0.34  & 20  & -3.99  \\
			0.74  & 7.75  & 62.00  & 0.34  & 19  & -3.99  \\
			0.74  & 7.75  & 62.00  & 0.33  & 18  & -3.98  \\
			0.74  & 7.74  & 62.00  & 0.33  & 17  & -3.97  \\
			0.74  & 7.73  & 62.00  & 0.33  & 16  & -3.96  \\
			0.74  & 7.72  & 62.00  & 0.33  & 15  & -3.96  \\
			0.74  & 7.70  & 62.00  & 0.33  & 14  & -3.95  \\
			0.74  & 7.67  & 62.00  & 0.33  & 13  & -3.94  \\
			0.74  & 7.59  & 62.00  & 0.33  & 12  & -3.92  \\
			0.74  & 7.35  & 62.00  & 0.33  & 11  & -3.89  \\
			0.74  & 7.00  & 62.00  & 0.33  & 10  & -3.84  \\
			0.74  & 6.63  & 62.00  & 0.33  & 9  & -3.77  \\
			0.74  & 6.25  & 62.00  & 0.33  & 8  & -3.69  \\
			0.74  & 5.88  & 62.00  & 0.33  & 7  & -3.61  \\
			0.74  & 5.51  & 62.00  & 0.33  & 6  & -3.54  \\
			0.74  & 5.14  & 62.00  & 0.33  & 5  & -3.48  \\
			0.74  & 3.70  & 62.00  & 0.33  & 4  & -3.43  \\
			0.74  & 3.69  & 62.00  & 0.33  & 3  & -3.43  \\ 
			0.74  & 3.68  & 62.00  & 0.33  & 2  & -3.43  \\ 
			\hline
		\end{tabular}
	\end{table}
	\FloatBarrier
	
	\begin{table}[ht]%
		\caption[Table]{Pareto optimal solutions calculated via MAiNGO for a upper bound on the input temperature of \SI{61}{\degreeCelsius}.}
		\label{tbl:data_MAINGO_61}
		\centering
		\begin{tabular}{|cccc|cc|} 
			\hline
			v\textsubscript{I} & v\textsubscript{M} & T & c\textsubscript{CTAB} & $\Delta$r$^2$\textsubscript{H} & F\textsubscript{product} \\ 
			
			[\SI{}{\milli\liter\per\minute}] & [\SI{}{\milli\liter\per\minute}] & [\SI{}{\degreeCelsius}] & [\SI{}{\milli\mole\per\liter}] & [\SI{}{\nano\meter\squared}] & [\SI{}{\milli\liter\per\minute}]\\
			
			\hline
			0.60  & 6.44  & 60.71  & 0.23  & 25  & -1.55  \\
			0.60  & 6.01  & 60.71  & 0.23  & 24  & -1.48  \\
			0.60  & 5.59  & 60.70  & 0.23  & 23  & -1.40  \\
			0.60  & 5.16  & 60.70  & 0.23  & 22  & -1.31  \\
			0.60  & 2.66  & 60.73  & 0.23  & 21  & -1.27  \\
			0.60  & 2.66  & 60.72  & 0.23  & 20  & -1.27  \\
			0.60  & 2.66  & 60.72  & 0.23  & 19  & -1.27  \\
			0.60  & 2.66  & 60.71  & 0.23  & 18  & -1.27  \\
			0.60  & 2.66  & 60.71  & 0.23  & 17  & -1.26  \\
			\hline
		\end{tabular}
	\end{table}
	\FloatBarrier